\documentclass[aps,pre,twocolumn,superscriptaddress,groupedaddress]{revtex4}  
\usepackage{graphicx}  
\usepackage{dcolumn}   
\usepackage{bm}        
\usepackage{amssymb}   
\usepackage{dsfont} 
\usepackage{physics}
\usepackage{lipsum}
\usepackage{mathtools}
\usepackage[export]{adjustbox}
\usepackage{hyperref}
\graphicspath{{./plots/}}

\newcommand{\ue}{\text{e}}

\usepackage{color}

\begin{document}

\title{Entanglement production by interaction quenches of quantum chaotic subsystems}
\author{Jethin J. Pulikkottil}
\affiliation{Department of Physics and Astronomy, Washington State University, Pullman, Washington 99164-2814. USA}
\author{Arul Lakshminarayan}
\affiliation{Department of Physics, Indian Institute of Technology Madras, Chennai 600036, India}
\affiliation{Max-Planck-Institut f{\"u}r Physik komplexer Systeme, N{\"o}thnitzer, Stra{\ss}e 38, 01187 Dresden, Germany.}
\author{Shashi C. L. Srivastava}
\affiliation{Variable Energy Cyclotron Centre, 1/AF Bidhannagar, Kolkata 700064, India}
\affiliation{Homi Bhabha National Institute, Training School Complex, Anushaktinagar, Mumbai - 400085, India}
\author{Arnd B{\"a}cker}
\affiliation{Technische Universit{\"a}t Dresden, Institut f{\"u}r Theoretische Physik and Center for Dynamics, 01062 Dresden, Germany}
\affiliation{Max-Planck-Institut f{\"u}r Physik komplexer Systeme, N{\"o}thnitzer, Stra{\ss}e 38, 01187 Dresden, Germany.}
\author{Steven Tomsovic}
\affiliation{Department of Physics and Astronomy, Washington State University, Pullman, Washington 99164-2814. USA}
\date{\today}

\begin{abstract}
The entanglement production in bipartite quantum systems
is studied for initially unentangled product eigenstates of the subsystems,
which are assumed to be quantum chaotic.
Based on a perturbative computation of the
Schmidt eigenvalues of the reduced density matrix,
explicit expressions for the time-dependence
of entanglement entropies, including the von Neumann entropy, are given.
An appropriate re-scaling of time and the entropies
by their saturation values leads a universal curve,
independent of the interaction.
The extension to the non-perturbative regime
is performed using a recursively embedded perturbation theory
to produce the full transition and the saturation values.
The analytical results are found to be in good agreement
with numerical results for random matrix computations
and a dynamical system given by a pair of coupled kicked rotors.
\end{abstract}

\maketitle

\section{Introduction}

Entanglement and its generation, besides its intrinsic interest as an unique quantum correlation and resource for quantum information processing \cite{HHHH2009, NieChu2010}, continues to be vigorously investigated due to its relevance to a wide range of questions such as
thermalization and the foundations of  statistical physics \cite{Popescu06,JC05,DeutchLiSharma2013,Kaufman16,Rigol2016}, decoherence \cite{Zurek91,Haroche1998,Zurek03,Davidovich_2016}, delocalization \cite{WSSL2008,kim13} and quantum chaos in few and many-body systems \cite{Miller99, Bandyopadhyay02, Wang2004,LS05,petitjean2006lyapunov, tmd08, Amico08,Chaudhary}. Some of these issues
concern the rate of entanglement production, the nature of multipartite entanglement sharing and distribution, and long-time saturation or indefinite growth. Details of entanglement production in integrable {\it versus} nonintegrable systems is an
active area of study, and it is generally appreciated that concomitant with the production of
near random states of nonintegrable systems is the production of large entanglement that can lead to thermalization
of subsystems.

Entanglement produced from suddenly joining two spin chains each in its ground state produces entanglement growth $\sim \ln(t/a)$ at long times~\cite{Calabrese_2007}, whereas quenches starting from arbitrary states can produce large entanglement including a linear growth phase \cite{JC05}. Similarly in an ergodic or eigenstate thermalized phase a system can show ballistic entanglement growth \cite{kim13,Ho17}, whereas in the many-body localized phase
it is known to have a logarithmic growth in time. The entanglement in almost all these studies relates to bipartite block entanglement between
macroscopically large subsystems in many-body systems.

Even earlier works have explored the entanglement in Floquet or periodically forced systems such as the coupled kicked tops or standard maps, as a means to study the relationship between chaos and entanglement \cite{FNP1998,Miller99,Lakshminarayan2001, Bandyopadhyay02,Fujisaki03,Bandyopadhyay04}.
It was seen that chaos in general increases bipartite entanglement and results in near
maximal entanglement as the states become typical, in the sense of Haar or uniform measure.
The initial states considered were mostly phase-space localized coherent states. In the case when the uncoupled systems are chaotic, and the interactions are weak, after a short Ehrenfest time scale
the growth of the entropy or entanglement is essentially as if the coherent states were initially random subsystem states.  In cases in which the subsystems are fully chaotic,
the growth of entanglement (beyond the Ehrenfest time) is dependent on the coupling strength rather than on measures of chaos such as the Kolmogorov-Sinai entropy or the Lyapunov exponent \cite{Fujisaki03,Bandyopadhyay04} . A linear growth was observed in this case and perturbation theory \cite{Fujisaki03} is successful in describing it, and the more extended time behavior was described by perturbation theory along with random matrix theory (RMT) \cite{Bandyopadhyay04}. The linear growth leads to saturation values that are interaction independent, and in cases of moderately large coupling this is just the bipartite entanglement of typical or random states in the full Hilbert space.

In sharp contrast is the case for which the initial states are eigenstates of the non-interacting fully chaotic systems. This presents a very different scenario for weak couplings that to our knowledge has not been previously studied. The present paper develops a full theory in this case, starting
from a properly regularized perturbation theory wherein a universal growth curve involving a suitably scaled time is derived. Importantly, this is further developed into a theory valid for non-perturbative strong couplings. We study these cases as an ensemble average over all uncoupled eigenstates, which clearly forms a special set of states in the Hilbert space for weak coupling.

 The entanglement production starts off linearly, as in the case of generic states, before saturating at much smaller entanglement values that are manifestly and strongly interaction dependent and reflect a ``memory" of the initial ensemble to which it belongs. Interestingly while the linear regime is independent of whether the system possesses time-reversal symmetry or not, the subsequent behavior including the saturation value of the entanglement is larger for the case when time-reversal is broken. For generic initial states time-reversal symmetry has not played a significant
 role in the entanglement production or saturation.
As the interaction is increased this saturation approaches that of random or typical values and then the memory of starting off as a special initial state no longer persists. An essential aspect of this study is to elucidate at what interaction strength such a transition happens.

The interaction is properly measured by a scaled dimensionless transition parameter $\Lambda$ that also determines transitions in the spectral statistics and eigenfunction entanglements of such systems \cite{Srivastava16,Lakshminarayan16,Tomsovic18c}. If $\Lambda=0$, the noninteracting case, although the two subsystems are chaotic, the quantum spectrum of the system has a Poisson level spacing statistic \cite{Tkocz12} and therefore has many nearly degenerate levels which start to mix when the subsystems are weakly coupled. If $\Lambda \ll 1$, we are in a perturbative regime wherein the eigenstates with appreciable entanglements have a Schmidt rank of approximately two, i.e.\ the reduced density matrix of the eigenstates has at most two principal nonzero eigenvalues. This circumstance carries over to time evolving states which are initially product eigenstates.  Universal features of the eigenstate entanglement depend only on the single scaled parameter $\Lambda$.  For example, the linear entropy of the eigenstates $\sim \sqrt{\Lambda}$ \cite{Lakshminarayan16,Tomsovic18c}.  On the other hand, as shown in this paper, time evolving states develop a linear entropy $\sim C(2,t)\sqrt{\Lambda}$, with $C(2,t)$ being a {\it universal} function for a properly scaled time, independent of the details of the interaction or the chaotic subsystem dynamics, except for a slight dependence on whether the system is time-reversal symmetric or not.

Such a universality follows from the existence of underlying RMT models that
describe the transition from uncoupled to strongly coupled systems, a RMT transition ensemble. Although it is standard to
apply RMT for stationary states and spectral statistics \cite{Bohigas84, Brody81, Haake10}, as indeed done for strongly chaotic and weakly
interacting systems \cite{Srivastava16,Lakshminarayan16,Tomsovic18c}, it is noteworthy that this is typically not valid for the time evolution, because RMT lacks correlations required for describing short time dynamics properly.  However the time scales
over which the entanglement develops is much longer than the Ehrenfest time after which specific dynamical system features disappear.  Thus, universal behaviors can be derived from such RMT transition ensembles provided the time scales of interest remain much longer than the Ehrenfest time scale. It turns out that for $\Lambda \gtrsim 1$, the interaction is strong enough that the
system has fluctuations that are typical of RMT of over the whole space, for example the consecutive neighbor spacing of eigenvalues is that of Wigner \cite{Srivastava16}. This also signals the regime for which eigenstates have typical entanglement of random states \cite{Lakshminarayan16,Tomsovic18c}  and as shown here, the time-evolving states lose memory of whether they initially belonged to special ensembles such as the noninteracting eigenstates.

Although regularized perturbation theory, initially developed for studying symmetry breaking in
statistical nuclear physics~\cite{French88a,Tomsovicthesis}, is used in the $\Lambda \ll 1$ regime, a novel recursive use of the perturbation theory allows for approximate, but very good extensions to the non-perturbative regime.  In fact, it covers well the full transition. This provides an impressive connection of the entanglement both as a function of time and as a function of the interaction strength to the RMT regime where nearly maximal entanglement is obtained and formulas such as Lubkin's for the linear entropy \cite{Lubkin78} and Page's for the von Neumann entropy \cite{Page93,Sen1996} are obtained.
We illustrate the general theory by specifically considering both time-reversal symmetric and violating RMT transition
ensembles, given respectively by subsystem Floquet operators chosen from the circular orthogonal ensemble (COE)
and the circular unitary ensemble (CUE) respectively. These are classic RMT ensembles consisting of unitary
matrices that are uniformly chosen with densities that are invariant under orthogonal (COE) and unitary (CUE) groups \cite{MehtaBook}.

In addition, we apply it to a dynamical system of coupled standard maps \cite{Froeschle71,Lakshminarayan2001,Richter14}. The standard map is a textbook example of a chaotic Hamiltonian system and is simply a periodically kicked pendulum. There is a natural translation symmetry in
the angular momentum that makes it possible to consider the classical map on a torus phase space, with periodic boundary
conditions in both position and momentum. This yields convenient finite dimensional models of quantum chaos, the dimension
of the Hilbert space being the inverse scaled Planck constant. The model we consider is that of coupling two
such maps which has proven useful in previous studies relating to entanglement \cite{Lakshminarayan2001,Lakshminarayan16,Tomsovic18c}, spectral transitions \cite{Srivastava16} and out-of-time-ordered correlators \cite{RaviLak2019}.

Two possible concrete examples are
a pair of particles in a chaotic quantum dot with tunable interactions or two spin chains that are in the ergodic phase before being suddenly joined.
Recent experiments have accessed information of time-evolving states of interacting few-body systems via
state tomography of single or few particles, facilitating the study of the role of entanglement in the
approach to thermalization of closed systems.
Specifically, experiments that have
studied nonintegrable systems include, for example, few qubit kicked top implementations \cite{Chaudhary,Neill2016} and the Bose-Hubbard Hamiltonian \cite{Kaufman16}. Thus modifications of these to accommodate weakly interacting parts are conceivable. This work adds to the already voluminous contemporary research on thermalization in closed systems
by looking in detail at the time evolution for a case when thermalization, in the sense of a typical subsystem entropy, is unlikely to occur \cite{GWEG2018}, namely starting from product eigenstates and quenching the interactions by suddenly turning them on.

This paper is organized as follows:
In Sec.~\ref{sec:background}
the necessary background material on entanglement in bipartite systems,
the random matrix transition ensemble and the unversal
transition parameter are given.
Sec.~\ref{sec:universal-entanglment-dynamics-perturbation-regime}
provides the perturbation theory for the
universal entanglement based on the eigenvalues of the reduced
density matrix, ensemble averaging for the CUE or COE
and invoking a regularization.
Based on this the eigenvalue moments of the reduced density matrix
are obtained in Sec.~\ref{sec:eigenvalue-moments}
leading to explicit expressions for the HCT entropies.
In particular it is shown that for small interaction
between the subsystems the simultaneous
re-scaling of time and of the entropies by their saturation
values leads to a universal curve which is independent
of the interaction.
The extension to the non-perturbative regime
is done in Sec.~\ref{sec:NonPerturbative}
by using a recursively embedded perturbation theory
to produce the full transition and the saturation values.
A comparison with a dynamical system given by
a pair of coupled kicked rotors is done
in Sec.~\ref{sec:coupled-kicked-rotors}.
Finally, a summary and outlook is given in
Sec.~\ref{sec:summary-and-outlook}.

\section{Background}
\label{sec:background}

\subsection{Entanglement in bipartite systems}

Consider pure states, $|\psi\rangle$, of a bipartite system whose Hilbert space is a tensor product space, $\mathcal{H}^A \otimes \mathcal{H}^B$, with subsystem dimensionalities, $N_A$ and $N_B$, respectively.  Without loss of generality,  let $N_A \leq N_B$.  The question to be studied is how much an initially unentangled state becomes entangled under evolution of some dynamics as a function of time.

The dynamics of a generic conservative system
could be governed by a Hamiltonian or by a unitary Floquet operator.
Specifically, a bipartite Hamiltonian system is of the form
\begin{equation}
H(\epsilon) = H_A \otimes \mathds{1}_B + \mathds{1}_A \otimes H_B + \epsilon V_{AB} \ , \label{eq:GenericHamiltonian}
\end{equation}
where the non-interacting limit is $\epsilon = 0$.  In the case of a quantum map, the dynamics can be described by a unitary Floquet operator~\cite{Srivastava16}
\begin{equation}
\mathcal{U}(\epsilon) = (U_A \otimes U_B) U_{AB}(\epsilon)\ ,
\label{eq:GenericFloquet}
\end{equation}
for which the non-interacting limit is $U_{AB}(\epsilon\rightarrow 0) = \mathds{1}$.  We assume that both $V_{AB}$ and $U_{AB}(\epsilon \neq 0)$ are entangling interaction operators~\cite{Lakshminarayan16}.

The Schmidt decomposition of a pure state is given by
\begin{equation}
  \ket{\psi} = \sum_{l=1}^{N_A} \sqrt{\lambda_l} \, \ket{\phi^A_l}\ket{\phi^B_l}.
  \label{eq:GenericSchmidtDecomposition}
\end{equation}
The normalization condition on the state $\ket{\psi}$ gives
\begin{equation} \label{eq:lambda-l-normalization}
  \sum_{l=1}^{N_A} \lambda_l = 1 .
\end{equation}
The state is unentangled if and only if the largest eigenvalue $\lambda_1 = 1$ (all others vanishing), and maximally entangled if $\lambda_l = 1/N_A$ for all $l$.  By partial traces, it follows that the reduced density matrices
\begin{equation}
\rho^A = \tr_B(\ket{\psi}\bra{\psi}), \qquad \rho^B = \tr_A(\ket{\psi}\bra{\psi})\ ,
\end{equation}
have the property
\begin{equation}
\rho^A \ket{\phi^A_l} = \lambda_l \ket{\phi^A_l}, \quad \text{and} \quad \rho^B \ket{\phi^B_l} = \lambda_l \ket{\phi^B_l},
\end{equation}
 respectively.  They are positive semi-definite, share the same non-vanishing (Schmidt) eigenvalues $\lambda_l$ and $\{ \ket{\phi^A_l} \}$, and $\{ \ket{\phi^B_l} \}$ form orthonormal basis sets in the respective Hilbert spaces.  For subsystem $B$ there are $N_B-N_A$ additional vanishing eigenvalues and associated eigenvectors.

A very useful class of entanglement measures are given by the von Neumann entropy and Havrda-Charv\'at-Tsallis (HCT) entropies~\cite{Bennett96,Havrda67,Tsallis88,Bengtsson06}.  The von Neumann entropy is given by
\begin{equation} \label{eq:GenericVonNeumannEntropy}
\begin{split}
S_1 & = -\tr_A(\rho^A \ln \rho^A ) = -\tr_B(\rho^B \ln \rho^B )\\
    &= -\sum_{l=1}^{N_A} \lambda_l \ln \lambda_l,
\end{split}
\end{equation}
which vanishes if the state is unentangled and is maximized if all the nonvanishing eigenvalues are equal to $1/N_A$.  The HCT entropies are obtained from moments of the Schmidt eigenvalues. Defining
\begin{equation}
\mu_\alpha = \tr_A[(\rho^A)^\alpha] = \tr_B[(\rho^B)^\alpha] = \sum_{l=1}^{N_A} \lambda_l^\alpha, \quad \alpha >0 \ ,
\label{eq:GenericMoments}
\end{equation}
gives the HCT entropies as
\begin{equation} \label{eq:GenericHCTEntropy}
  S_\alpha = \frac{1-\mu_\alpha}{\alpha-1}.
\end{equation}
Note that these differ from the R\'enyi entropies \cite{Ren1961wcrossref},
which are defined by
\begin{equation*}
  R_\alpha = \dfrac{\ln \mu_{\alpha}}{1-\alpha} .
\end{equation*}
In the limit $\alpha \to 1$ also $R_\alpha$ turns into the
von Neumann entropy.
In this work we use the HCT entropies as performing ensemble averages
is easier using $\mu_{\alpha}$ than $\ln \mu_{\alpha}$.

\subsection{Quantum chaos, random matrix theory, and universality}

Many statistical properties of strongly chaotic quantum systems are successfully modeled and derived with the use of RMT~\cite{Brody81,Bohigas84}.  Generally speaking, the resulting properties are universal, and in particular, do not depend on any of the physical details of the system with the exception of symmetries that it respects.  Here the subsystems are assumed individually to be strongly chaotic.  Thus, the statistical properties of the dynamics, Eq.~(\ref{eq:GenericFloquet}),  can be modeled with the operators $U_A$ and $U_B$ being one of the standard circular RMT ensembles~\cite{Dyson62e}, orthogonal, unitary, or symplectic, depending on the fundamental symmetries of the system~\cite{Porterbook}.

We concentrate on the orthogonal (COE) and unitary ensembles (CUE) depending on whether or not time reversal invariance is preserved, respectively.
The derivation of the typical entanglement production for some initial state relies on the dynamics governed by
the random matrix transition ensemble \cite{Srivastava16,Lakshminarayan16}
\begin{equation}
\mathcal{U}_{\text{RMT}}(\epsilon) = (U_A^\text{RMT} \otimes U_B^\text{RMT}) U_{AB}(\epsilon)\ .
\label{eq:GenericFloquetRMT}
\end{equation}
The operator $U_{AB}(\epsilon)$ is assumed to be diagonal in the direct product basis of the two subsystem ensembles. Explicitly, the diagonal elements are considered to be of the form $\exp(2 \pi i \epsilon  \xi_{kl})$, where $\xi_{kl}$ ($1\le k,l\le N_A,N_B$) is a random number uniformly distributed in $(-1/2,1/2]$.

\subsection{Symmetry breaking and the transition parameter}

The statistical properties of weakly interacting quantum chaotic bipartite systems have been studied recently, with the focus on spectral statistics, eigenstate entanglement, and measures of localization~\cite{Srivastava16,Lakshminarayan16,Tomsovic18c}.  If the subsystems are not interacting, the spectrum of the full system is just the convolution of the two subsystem spectra giving an uncorrelated spectrum in the large dimensionality limit.  The eigenstates of the system are unentangled.  It is very fruitful to conceptualize this as a dynamical symmetry.  Upon introducing a weak interaction between the subsystems, this symmetry is weakly broken.  As the interaction strength increases, the spectrum becomes increasingly correlated, and the eigenstates entangled.

Here $U_{AB}(\epsilon)$ plays the role of a dynamical symmetry breaking operator.  For $\epsilon=0$, the symmetry is preserved (the dynamics of the subsystems are completely independent), and as $\epsilon$ gets larger, the more complete the symmetry is broken.  It is known that for sufficiently chaotic systems, there is a universal scaling given by a transition parameter which governs the influence of the symmetry breaking on the system's statistical properties~\cite{French88a}.  The transition parameter is defined as~\cite{Pandey83}
\begin{equation}
\Lambda = \frac{v^2(\epsilon)}{D^2},  \label{eq:LambdaDef}
\end{equation}
where $D$ is the mean level spacing and $v^2(\epsilon)$ is the mean square matrix element in the eigenbasis of the symmetry preserving system, calculated locally in the spectrum.

For the COE and CUE the leading behavior in $N_A$ and $N_B$ is \cite{Srivastava16,Tomsovic18c,HerKieFriBae2019:p}
\begin{align}
\label{eq:Lambda-RMT}
& \, \Lambda = \frac{N_A N_B}{4 \pi^2} \left[1-\dfrac{\sin^2 (\pi \epsilon)}{\pi^2\epsilon^2} \right]
\sim \frac{\epsilon^2 N_AN_B}{12}\ ,
\end{align}
where the last result is in the limit of large $N_A$, $N_B$.
The transition parameter $\Lambda$ ranges over  $0 \le \Lambda \le N_A N_B/4\pi^2\ (N_A, N_B  \rightarrow \infty)$, where the limiting cases are fully symmetry preserving, and fully broken, respectively.
In essence, the latter expression of Eq.~\eqref{eq:Lambda-RMT}  illustrates the fact that as the system size grows, a symmetry breaking transition has a discontinuously fast limit in $\epsilon$.

The transition parameter gives the relation necessary to compare the statistical properties of systems of any size and kind to each other.  As long as $\Lambda$ has identical values, the systems have identical properties.  However, for a particular dynamical system, it can turn out to be rather difficult to calculate $\Lambda$.  Although, the statistical properties are universal and independent of the nature of the system in this chaotic limit, properties such as whether the system is many-body or single particle, Fermionic or Bosonic, actually enter into its calculation.  For example, a method for calculating  $\Lambda$ for highly excited heavy nuclei is given in Ref.~\cite{French88b}.  The far simpler case of coupled kicked rotors is given in Ref.~\cite{Srivastava16}, and is used ahead for illustration.  In extended systems, the issue of localization emerges, which must also be taken into account, and for them the term sufficiently chaotic is meant to exclude a localized regime.

\section{Universal entanglement production -- Perturbative regime}
\label{sec:universal-entanglment-dynamics-perturbation-regime}

The starting point of a derivation of the typical production rate of entanglement in initially unentangled states is the random matrix transition ensemble~\eqref{eq:GenericFloquetRMT}.  Following a similar derivation sequence for the eigenstates in Refs.~\cite{Lakshminarayan16,Tomsovic18c}, the first step is to derive expressions for the eigenvalues of the reduced density matrix, which can be obtained from the Schmidt decomposition of the time evolved state of the system. Applying a standard Rayleigh-Schr{\"o}dinger perturbation theory leads to perturbation expressions for the Schmidt eigenvalues.  However, due to the Poissonian fluctuations in the spectrum of the non-interacting system, near-degeneracies occur too frequently and cause divergences in the ensemble averages. It is therefore necessary first to regularize the eigenvalue expressions appropriately.  It also turns out that the perturbation expressions for the HCT entanglement measures can be further extended to a non-perturbative regime by recursively invoking the regularized perturbation theory leading to a differential equation, which is analytically solvable~\cite{Tomsovic18c},
see Sec.~\ref{sec:NonPerturbative}.

\subsection{Definitions}

The eigenvalues and corresponding eigenstates of the unitary operators
$U_A$ and $U_B$ for the subsystems
and of $\mathcal{U}(\epsilon) = (U_A \otimes U_B) U_{AB}(\epsilon)$
of the full bipartite system \eqref{eq:GenericFloquet} are given
by the equations
\begin{eqnarray}
U_A \ket{j^A} &=& \ue^{i \theta_j^A}  \ket{j^A},\quad j=1,2,3,\ldots,N_A \nonumber \\
U_B \ket{k^B} &=& \ue^{i \theta_k^B} \ket{k^B}, \quad k=1,2,3,\ldots,N_B\\
\mathcal{U}(\epsilon) \ket{\Phi_{jk}} &=& \ue^{i \varphi_{jk}} \ket{\Phi_{jk}}\ .
\nonumber
\end{eqnarray}
To simplify the notation, the superscripts $A$ and $B$ are dropped for both eigenkets, $\ket{j^A}\ket{k^B} \equiv \ket {jk}$, and the eigenvalues $\theta_j^A \equiv \theta_j$ ($\theta_k^B \equiv \theta_k$). It is understood that the labels $j$ and $k$ are reserved for the subsystems $A$ and $B$, respectively. Similarly for convenience, the subscript $AB$ is dropped from the operator $V_{AB}$.

Given the form \eqref{eq:GenericFloquet} of the unitary operator $\mathcal{U}(\epsilon)$, in the limit $\epsilon \rightarrow 0$ one has $\ket{\Phi_{jk}} \rightarrow \ket{jk}$ which is a product eigenstate of the unperturbed system and forms a complete basis with spectrum $\varphi_{jk} \rightarrow \theta_{jk} = \theta_j + \theta_k \,\,\text{mod}\,\, 2\pi$.  For non-vanishing $\epsilon$ there is a unitary transformation $S$ between the eigenbases for the set $\ket{\Phi_{jk}}$ and $\ket{jk}$ whose matrix elements can be identified using the relations
\begin{eqnarray}
\ket{\Phi_{jk}} &=& \sum_{j'k'} S_{jk,j'k'} \ket{j'k'} = \sum_{j'k'}  \ket{j'k'} \bra{j'k'}\ket{\Phi_{jk}} \nonumber \\
\ket{jk} &=& \sum_{j'k'} S^\dagger_{jk,j'k'} \ket{\Phi_{j'k'}}\ .
\label{eq:UnitaryTransform}
\end{eqnarray}

\subsection{Eigenvalues of the reduced density matrix}

In the limit $N_A \rightarrow \infty$, perturbation theory for unitary Floquet systems generates the same equations as for Hamiltonian systems up to vanishing corrections of $\mathcal{O}((N_A N_B)^{-1})$ if one identifies $U_{AB}(\epsilon)=\exp(i\epsilon V)$~\cite{Tomsovic18c}.  For an initial unentangled state, begin by considering an eigenstate $\ket{jk}$ of the non-interacting system.  Denote the time evolution of this initial state after $n$ iterations of the dynamics as $\ket{j k(n;\epsilon)}$ [$ = \mathcal{U}^n(\epsilon) \ket{j k}$].  Upon the usual insertion of the completeness relation one gets
\begin{equation}
\ket{jk(n;\epsilon)} =  \sum_{j'k'} \ue^{i n \varphi_{j'k'}} S^\dagger_{jk,j'k'}  \ket{\Phi_{j'k'}}. \label{eq:GeneralTimeEvolvedState}
\end{equation}
This time evolved state has a standard Schmidt decomposed form
\begin{equation}
\ket{jk(n;\epsilon)} = \sum_{l = 1}^{N_A} \sqrt{\lambda_l(n;\epsilon)} \, \ket{\phi^A_l(n;\epsilon)}\ket{\phi^B_l(n;\epsilon)},
\label{eq:GeneralTimeEvolvedStateInSDForm}
\end{equation}
where $\lambda_1 \geq \lambda_2 \geq \ldots \geq \lambda_{N_A}$ are time-dependent Schmidt numbers (eigenvalues of the reduced density matrices) such that $\sum_l \lambda_l(n;\epsilon) = 1$, and $\{ \ket{\phi^A_l(n;\epsilon)}\}$, $\{ \ket{\phi^B_l(n;\epsilon)}\}$ are the corresponding Schmidt eigenvectors of the $A$ and $B$ subspaces, respectively.

It was shown in Ref.~\cite{Lakshminarayan16} that for weak perturbations, the Schmidt decomposition of the eigenstates to $\mathcal{O}(N_A^{-1})$ corrections are given by the neighboring eigenstates of the unperturbed (non-interacting) system and the perturbation theory coefficients.  This can be considered as a kind of automatic Schmidt decomposition.
 The generalization to the time evolving states, $\ket{j k(n;\epsilon)}$, follows by another insertion of the unitary transformation $S$ to give
\begin{equation}
\ket{jk(n;\epsilon)} = \sum_{j''k''} \sum_{j'k'} \ue^{i n \varphi_{j'k'}} S^\dagger_{jk,j'k'} S_{j'k',j''k''} \ket{j''k''}. \label{eq:GeneralTimeEvolvedState2}
\end{equation}
This leads to the identification
\begin{equation}
\lambda_l(n;\epsilon) = \left| \sum_{j'k'} \ue^{i n \varphi_{j'k'}} S^\dagger_{jk,j'k'} S_{j'k',(jk)_l} \right|^2,\label{eq:SchmidtNumbersPT}
\end{equation}
where $j''k''\rightarrow (jk)_l$, meaning that fixing $l$ fixes a unique and distinct index pair $(jk)_l$; e.g.~$(jk)_1=jk$.  Only a small subset ($\lesssim N_A$) of possible pairs $j''k''$ are related to a $(jk)_l$ due to the energy denominators in perturbation theory.  This is a direct result of the automatic Schmidt decomposition.
For $\epsilon = 0$ one has $\lambda_1(n;0) =1$, and the rest of the Schmidt eigenvalues vanish
by the normalization \eqref{eq:lambda-l-normalization}, as the initial state
is a product state of eigenstates of the two subsystems.

To prepare for ensemble averaging, it is helpful to: i) assume that the $jk$ pairs are ordered by the order of the eigenvalues $\varphi_{jk}$, ii) use the properties of $S$ so that the $n=0$ results are immediately evident, and iii) separate out the diagonal matrix element $S_{jk,jk}$ as a special case.  Let $\Delta \varphi_{jk,j^\prime k^\prime} = \varphi_{jk} - \varphi_{j^\prime k^\prime}$. For the largest eigenvalue, i.e.~Eq.~(\ref{eq:SchmidtNumbersPT}) for $l = 1$, one finds
\begin{align}
& \lambda_1(n;\epsilon) =  1 - 2  \sum_{j'k', j''k''} \left|S_{j'k',jk}\right|^2 \left|S_{j''k'',jk}\right|^2 \nonumber \\
& \qquad \qquad \qquad \qquad \qquad \quad \times \sin^2 \left(\frac{n \Delta \varphi_{j'k',j''k''}}{2} \right) \nonumber \\
& = 1 - 4 \left|S_{jk,jk}\right|^2\sum_{j'k'}  \left|S_{j'k',jk}\right|^2  \sin^2 \left(\frac{n \Delta \varphi_{jk,j'k'}}{2} \right) \nonumber \\
& - 4  \sum_{\substack{j'k' \le j''k'' \\ \ne jk}} \left|S_{j'k',jk}\right|^2 \left|S_{j''k'',jk}\right|^2 \sin^2 \left(\frac{n \Delta \varphi_{j'k',j''k''}}{2} \right).  \nonumber \\
\label{eq:LargestEigenvalueRaw}
\end{align}
For $l \ge 2$, and thus $(jk)_l \neq jk$, a similar manipulation gives
\begin{align}
&\lambda_l(n;\epsilon) = - \sum_{\substack{j'k'\\j''k''}} \Re\left\{S^\dagger_{jk,j'k'} S_{j'k',(jk)_l} S^\dagger_{(jk)_l,j''k''} S_{j''k'',jk} \right\}  \nonumber \\
&  \qquad \qquad \qquad \qquad \times 2 \sin^2 \left(\frac{n \Delta \varphi_{j'k',j''k''}}{2} \right) \nonumber \\
& \qquad \qquad - \sum_{\substack{j'k'\\j''k''}} \Im\Big\{S^\dagger_{jk,j'k'} S_{j'k',(jk)_l} S^\dagger_{(jk)_l,j''k''} S_{j''k'',jk} \Big\} \nonumber \\
& \qquad \qquad \qquad \qquad \times \sin \left(n \Delta \varphi_{j'k',j''k''} \right).
\label{eq:OtherEigenvaluesRaw}
\end{align}
Note that summing Eq.~\eqref{eq:OtherEigenvaluesRaw}
over $l>1$ reproduces unity minus the expression of Eq.~\eqref{eq:LargestEigenvalueRaw} as it must.  These Schmidt eigenvalue expressions are exact to order $\mathcal{O}(N_A^{-1})$.

Lowest order Rayleigh-Schr\"{o}dinger perturbation theory is applied to the matrix elements of $S$ in order to obtain the complete $O(\epsilon^2)$ terms of the corresponding expressions for the Schmidt eigenvalues.  Let $\Delta \theta_{jk,j'k'} = \theta_{jk}-\theta_{j'k'}$. The matrix elements $S_{jk,j'k'}$ are approximately
\begin{equation}
S_{jk,j'k'} \approx
\begin{cases}
\frac{1}{\sqrt{\mathcal{N}_{jk}}}, &  jk = j'k' \\
\frac{1}{\sqrt{\mathcal{N}_{jk}}} \frac{\epsilon \, V_{j'k',jk}}{\Delta \theta_{jk,j'k'}},   & jk \neq j'k',
\end{cases} \label{eq:UnitaryTransformPT}
\end{equation}
where $\mathcal{N}_{jk}$ is the normalization factor and the perturbed quasienergy is
\begin{align}
& \varphi_{jk} = \theta_{jk} +  \epsilon^2 \sum_{j'k' \neq jk } \frac{|V_{j'k',jk}|^2}{\Delta \theta_{jk,j'k'}}. \label{eq:Unreg_quasienergy}
\end{align}
Note first that in this derivation the diagonal matrix elements $V_{j'k',j'k'}$ are set to zero, because the energy shift due to the first order correction is a random number added to an uncorrelated spectrum giving another uncorrelated spectrum, and hence will not change the spectral statistics nor rotate the eigenvectors.  Secondly, the normalization factor is included even though its first correction is $O(\epsilon^2)$ because it plays a significant role in determining the regularized expressions ahead, likewise for the perturbed eigenvalues multiplied by the time in the argument of the sine function.  For the largest eigenvalue Eq.~(\ref{eq:LargestEigenvalueRaw}) becomes
\begin{align}
&\lambda_1(n;\epsilon) \approx 1 - \frac{4}{\mathcal{N}_{jk}} \sum_{j'k' \neq jk} \bigg(\frac{\epsilon^2 \, |V_{jk,j'k'}|^2}{\mathcal{N}_{j'k'} \, \Delta \theta_{j'k',jk}^2} \bigg)  \nonumber \\
& \qquad \qquad \qquad \qquad \times \sin^2 \left(\frac{n \Delta \varphi_{j'k',jk}}{2}\right),
\label{eq:LargestEigenvaluePT}
\end{align}
and for $l \ge 2$ the others, Eq.~\eqref{eq:OtherEigenvaluesRaw}, read
\begin{align}
& \lambda_l(n;\epsilon) \approx \frac{4}{\mathcal{N}_{jk}}
    \frac{\epsilon^2\,|V_{(jk)_l,jk}|^2}{\mathcal{N}_{(jk)_l}\,\Delta \theta_{jk,(jk)_l}^2} \sin^2 \left(\frac{n \Delta \varphi_{(jk)_l,jk}}{2}\right),
\label{eq:OtherEigenvaluesPT}
\end{align}
where $\mathcal{N}_{jk}=\mathcal{N}_{(jk)_l}=1 + \mathcal{O}(\epsilon^2)$.

\subsection{Ensemble averaging}
Before moving on to ensemble averaging, it is helpful to make some rescalings as follows:
\begin{align}
& \Delta \theta_{j'k',jk} = D \, s_{j'k'}\\
& \Delta \varphi_{j'k',jk} = D \, s_{j'k'}(\epsilon) \approx D s_{j'k'} \bigg( 1 + \frac{2 \Lambda w_{j'k'}}{s^2_{j'k'}} \bigg) \label{eq:UnregSpacing}\\
& \epsilon^2 |V_{jk,j'k'}|^2 = \Lambda D^2 \, w_{j'k'}\ ,
\end{align}
where $D = 2\pi/(N_A N_B)$ is the mean level spacing of the full system, $s_{j'k'}=s_{j'k'}(0)$, and the $jk$ subscript is dropped where unnecessary.  The approximation in Eq.~\eqref{eq:UnregSpacing} follows by considering only the matrix element that directly connects the two levels.  The other terms in Eq.~\eqref{eq:UnregSpacing} move the levels back and forth and mostly cancel, but this term pushes the two levels away from each other and is dominant when the two levels are close lying where the correction may contribute.
The symmetry breaking (entangling) interaction matrix elements of $V$ represented in the eigenbasis of the unperturbed system behave as complex Gaussian random variable such that
$ \overline{ | V_{jk,j'k'} |^2 } = v^2 w_{j'k'} $, where $w_{j'k'}$ follow a Porter-Thomas distribution \cite{PorTho1956} for the COE and an exponential one for the CUE:
\begin{equation}
\rho(w) =
\begin{cases}
\frac{1}{\sqrt{2\pi w}}\exp(-w/2) & \qquad \text{for COE} \\
\exp(-w)                          & \qquad \text{for CUE}.
\end{cases}\label{PorterThomas}
\end{equation}
In both of the cases, $\overline{w_{j'k'} }= 1$, which is consistent with $\Lambda = \epsilon^2 v^2 /D^2$.  In real dynamical systems, deviations from Porter-Thomas distributions may occur as noted in Ref.~\cite{Tomsovic18c}.

Thus, in the rescaled variables the Schmidt eigenvalues for $l \ge 2$ are
\begin{equation}
\lambda_l(n;\Lambda) \approx \frac{4}{\mathcal{N}_{jk}}\Bigg(\frac{\Lambda \,w_{(jk)_l} }{\mathcal{N}_{(jk)_l}\, s_{(jk)_l}^2} \Bigg) \,\sin^2\bigg(\frac{n Ds_{(jk)_l}{(\epsilon)}}{2}\bigg), \label{eq:OtherEigenvaluesRescaled}
\end{equation}
and the relation, following from the normalization
condition Eq.~\eqref{eq:lambda-l-normalization},
\begin{equation}
\lambda_1(n;\Lambda) = 1 - \sum_{l\ne 1}\lambda_l(n;\Lambda)
\end{equation}
is exactly preserved to this order.  Next convert the expressions for the Schmidt eigenvalues into integrals, by making use of the function $R(s,w)$~\cite{Tomsovic18c},
\begin{equation}
R(s,w) = \sum_{j'k' \neq jk} \delta(w-w_{j'k'}) \delta(s-s_{j'k'}), \label{eq:R-function}
\end{equation}
which after ensemble averaging becomes the joint probability density of finding a level at a rescaled distance $s$ from $\theta_{jk}$ and the corresponding scaled matrix element $w_{j'k'}$ at the value $w$.
With these definitions, scalings, and substitutions, Eq.~(\ref{eq:LargestEigenvaluePT}) becomes
\begin{align}
& \lambda_1(n;\epsilon) \approx 1 - 4 \Lambda \int_{-\infty}^\infty \dd{s} \int_0^\infty \dd{w} \frac{w}{s^2}\,R(s,w)  \nonumber \\
&\qquad \qquad \qquad \times \sin^2\Big(\frac{n D s}{2}\big[\,1+\frac{2 \Lambda w}{s^2}\,\big]\Big).
\end{align}
The ensemble average of $\lambda_1(n;\epsilon)$ follows by substituting the ensemble average of $R(s,w)$ by
\begin{equation}
\overline{R(s,w)} = R_2(s) \, \rho(w),
\end{equation}
where $R_2(s)$ is the two-point correlation function and $\rho(w)$ is defined in Eq.~(\ref{PorterThomas}).   For an uncorrelated spectrum $R_2(s)=1$ for $-\infty < s < \infty$.
Therefore, the averaged largest Schmidt eigenvalue is
\begin{align}
& \overline{ \lambda_1(n;\Lambda)}\approx 1 - 4 \Lambda \int_{-\infty}^\infty \dd{s} \int_0^\infty \dd{w} \frac{w}{s^2}\, \rho(w) \nonumber \\
&\qquad \qquad \qquad \times \sin^2\Big(\frac{n D s}{2}\big[\,1+\frac{2 \Lambda w}{s^2}\,\big]\Big). \label{eq:LargestEigenvalueUnreg}
\end{align}
This expression diverges due to the fact that too many spacings are vanishingly small across the ensemble, and the perturbation theory must account for spacings smaller than the matrix elements.  In the next subsection the expressions are regularized properly for small $s$.

It is worth noting that if the interest is in the ensemble average of some function of $\lambda_1(n;\Lambda)$, then one must consider the ensemble average of the same function of $R(s,w)$.  Perhaps, the simplest example is the ensemble average of the square of  $\lambda_1(n;\Lambda)$ for which the needed result is~\cite{Tomsovic18c}
\begin{eqnarray}
\overline{R(s_1,w_1)R(s_2,w_2)} &=& R_3(s_1,s_2) \rho(w_1)\rho(w_2) + \nonumber \\
&&\hspace*{-2cm}
   \delta\left(w_1-w_2\right)\rho(w_1) \delta\left(s_1-s_2\right)R_2(s_1)
\label{threepoint}
\end{eqnarray}
which involves both the $2$-point and $3$-point spectral correlation functions.  However, it turns out that the leading correction depends on $R_2(s)$, as the $R_3(s_1,s_2)$ term gives a contribution that is $\mathcal{O}(\sqrt{\Lambda})$ smaller in comparison, and for example, generating the leading correction of high order moments depends only on the $2$-point spectral correlation function.  This circumstance is helpful ahead in the next section.

Following the same sequence of steps for the second largest eigenvalue $\lambda_2$ requires, in addition, the probability density of the closest scaled energy of one of the $\ket{(jk)_l}$.  For uncorrelated spectra it is given by $\rho_{\text{CN}}(s) = 2 \exp(-2 s)$ for $0 \le s < \infty$~\cite{Tomsovic18c,Srivastava19}.  One finds for the ensemble average of second largest eigenvalue,
\begin{align}
& \overline{\lambda_2(n;\Lambda)} \approx 4 \Lambda \int_{-\infty}^\infty \dd{s} \int_0^\infty \dd{w} \frac{w}{s^2} \,\rho(w) \, \rho_{\text{CN}}(s)\nonumber \\
&\qquad \qquad \qquad \times \sin^2\Big(\frac{n D s}{2}\big[\,1+\frac{2 \Lambda w}{s^2}\,\big]\Big),\label{eq:OtherEigenvalueUnreg}
\end{align}
which is also divergent for small $s$.  It turns out that the apparent order of corrections, $\mathcal{O}(\Lambda)$, seen in Eqs.~(\ref{eq:LargestEigenvalueUnreg}, \ref{eq:OtherEigenvalueUnreg}), is not correct.  The regularization required to deal with the small energy denominators in the perturbation expressions alters the leading order to $\mathcal{O}(\sqrt{\Lambda})$.

\subsection{Regularized perturbation theory}

The method for regularizing the perturbation expressions was introduced in Refs.~\cite{Tomsovicthesis,French88a} and developed for the Schmidt eigenvalues pertaining to the eigenstates of interacting quantum chaotic systems in Refs.~\cite{Lakshminarayan16,Tomsovic18c}. The standard Rayleigh-Schr\"{o}dinger perturbation expressions break down when the unperturbed spectrum has nearly degenerate levels due to the small energy denominators.  However, there is an infinite sub-series of terms within the perturbation series of a quantity of interest which involve only two levels that are diverging due to near-degeneracy. This subseries can be resummed to get the corresponding regularized expressions.  These results are equivalent to the two-dimensional degenerate perturbation theory results.

The regularized expressions for the Schmidt eigenvalues upon resummation of the two-level like terms of the perturbation series boil down to essentially replacing
\begin{align}
\frac{1}{\mathcal{N}_{jk}}\Bigg(\frac{\Lambda\,w_{j'k'} }{\mathcal{N}_{j'k'}\, s_{j'k'}^2} \Bigg) \mapsto \frac{\Lambda \, w_{j'k'}}{s_{j'k'}^2+ 4 \, \Lambda \, w_{j'k'}} , \label{eq:RegularizedNormTimesMatrixElement}
\end{align}
along with the energy spacing~\cite{Srivastava16} in Eq.~(\ref{eq:UnregSpacing}) as
\begin{equation}
s_{j'k'}(\epsilon)  \mapsto \sqrt{s_{j'k'}^2 + 4 \Lambda w_{j'k'} } \label{eq:RegularizedSpacing}
\end{equation}
in Eqs.~(\ref{eq:LargestEigenvaluePT}, \ref{eq:OtherEigenvaluesPT}).  To verify the result in Eq.~(\ref{eq:RegularizedNormTimesMatrixElement}), the Schmidt eigenvalues in Eqs.~(\ref{eq:LargestEigenvalueRaw}, \ref{eq:OtherEigenvaluesRaw}) were expanded using perturbation theory of the matrix elements $S_{jk,j'k'}$ up to and including order $\mathcal{O}(\epsilon^4)$.  The details for this are given in App.~\ref{app:RegularizationDerivation}.  For a two-level system the normalization is
\begin{equation}
|S_{jk,jk}|^2 = \frac{1}{\mathcal{N}_{jk}} = \frac{1}{2}\Bigg( 1 + \frac{|s_{j'k'}|}{\sqrt{s_{j'k'}^2+4 \Lambda w_{j'k'}}} \Bigg)
\end{equation}
and the matrix element
\begin{equation}
|S_{j'k',jk}|^2 = \frac{1}{2}\Bigg( 1 - \frac{|s_{j'k'}|}{\sqrt{s_{j'k'}^2+4 \Lambda w_{j'k'}}} \Bigg).
\end{equation}
Using these results, we get Eq.~(\ref{eq:RegularizedNormTimesMatrixElement}). This gives the regularized Schmidt eigenvalues for $l \neq 1$ as
\begin{align}
& \lambda_l(n;\Lambda) = \frac{4 \, \Lambda \, w_{(jk)_l}}{s^2_{(jk)_l} + 4 \, \Lambda \, w_{(jk)_l} } \nonumber \\
& \qquad \qquad \qquad \times \sin^2 \Big( \frac{n D}{2} \sqrt{s^2_{(jk)_l} + 4 \, \Lambda \, w_{(jk)_l}} \,\Big). \label{eq:SchmidtEigenvaluelneq1Reg}
\end{align}
Rescaling the spacing $z = s/\sqrt{\Lambda}$ and time
\begin{equation}  \label{eq:time-rescaling}
   t = n D\sqrt{\Lambda} ,
\end{equation}
the ensemble average of the first two Schmidt eigenvalues is given by
\begin{align}
& \overline{ \lambda_1(t;\Lambda)} = 1- \sqrt{\Lambda}\int_0^\infty \dd{w} \int_{-\infty}^\infty \dd{z} \frac{4 w}{z^2 + 4 w} \, \rho(w)  \nonumber \\
& \qquad \qquad \qquad \times  \sin^2\Big(\frac{t}{2} \sqrt{z^2 + 4 w} \Big) \nonumber \\
& \qquad \quad \, = 1 - C_2(1;t)\,\sqrt{\Lambda},
\end{align}
where $C_2(1;t)$ is a short-hand for the integral
(the general notation for arbitrary moments is given in
Eq.~(\ref{eq:C2Integral}) ahead) and,
\begin{align}
& \overline{ \lambda_2(t;\Lambda)} = \sqrt{\Lambda}\int_0^\infty \dd{w} \int_0^\infty \dd{z} \frac{4 w}{z^2 + 4 w} \nonumber \\
& \qquad \qquad \qquad \times  \, \rho(w) \,\big(2 \,\ue^{-2 z \sqrt{\Lambda}}\,\big) \sin^2\Big(\frac{t}{2} \sqrt{z^2 + 4 w} \Big) \nonumber \\
& \, \quad \qquad = C_{2}(1;t) \sqrt{\Lambda} + \mathcal{O}(\Lambda \ln \Lambda).
\end{align}
For sufficiently small $\Lambda$ it turns out that for both the COE and CUE cases, only (unperturbed) eigenstates corresponding to the first two largest Schmidt eigenvalues contribute largely to the state $\ket{jk(t;\Lambda)}$, i.e.,
\begin{equation}
\overline{ \lambda_1(t;\Lambda) + \lambda_2(t;\Lambda)} = 1 + \mathcal{O}(\Lambda \ln \Lambda ),
\end{equation}
and other Schmidt eigenvalues ($l>2$) contribute in higher orders than $\sqrt{\Lambda}$.
This is crucial for extending the perturbation theory of the Schmidt eigenvalue moments to the non-perturbative regime, which is done in Sec.~\ref{sec:NonPerturbative}.
It should be noted that as the unperturbed spectrum is uncorrelated, there is a non-zero probability of three-level, four-level and so-forth near-degeneracy occurrences, but with lower probability from the two-level case, and hence their contributions are higher of order than $\sqrt{\Lambda}$.

Moreover note that the perturbation expressions for the Schmidt eigenvalues of the eigenstates $\{\ket{\Phi_{jk}}\}$ given in Refs.~\cite{Lakshminarayan16, Tomsovic18c} for the largest eigenvalue and the other eigenvalues are $1 -  \sum_{j'k \neq jk} \epsilon^2 |V_{(j'k',jk}|^2/\Delta\theta^2_{jk,j'k'}$ and $\epsilon^2 |V_{(jk)_l,jk}|^2/\Delta\theta^2_{jk,(jk)_l}$, respectively, in contrast to the expressions for the Schmidt eigenvalues of a time evolving state $\ket{jk(n;\epsilon)}$ presented in Eqs.~(\ref{eq:LargestEigenvaluePT}, \ref{eq:OtherEigenvaluesPT}). Due to an extra normalization factor in the denominators of Eqs.~(\ref{eq:LargestEigenvaluePT}, \ref{eq:OtherEigenvaluesPT}), the expression for the regularization, although related, takes on a different form than that in Refs.~\cite{Lakshminarayan16, Tomsovic18c}.

\section{Eigenvalue moments of the reduced density matrix}
\label{sec:eigenvalue-moments}

To fully characterize the entanglement of the evolving state, the Schmidt eigenvalue expression in Eq.~(\ref{eq:SchmidtEigenvaluelneq1Reg}) is used to compute the leading order of general moments analytically and thereby the HCT entropies, good up to and including $\mathcal{O}(\sqrt{\Lambda})$.

\subsection{General moments} \label{subsec:GeneralMomentsCalc}

\begin{figure}
\includegraphics[width=8.6cm]{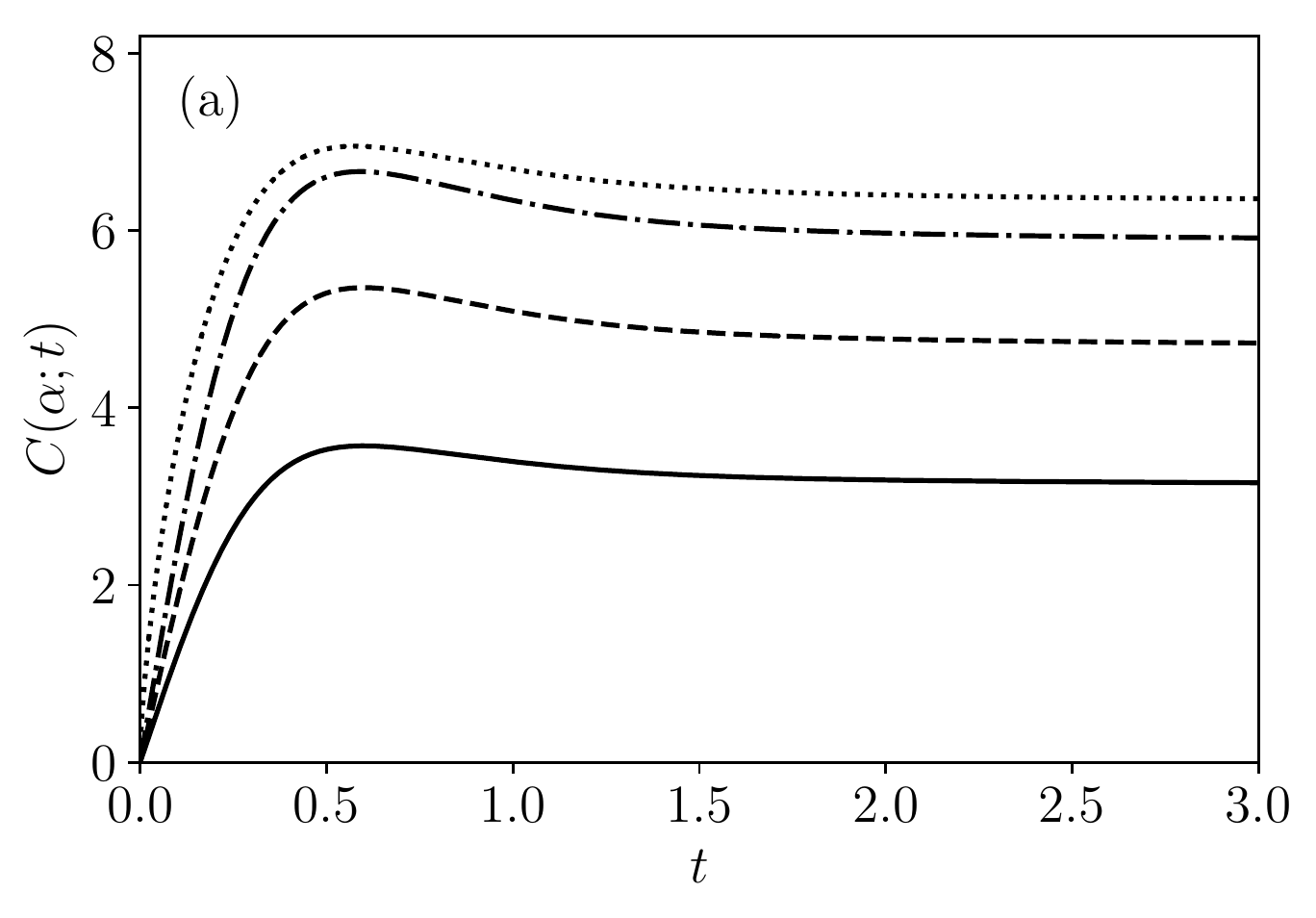}

\includegraphics[width=8.6cm]{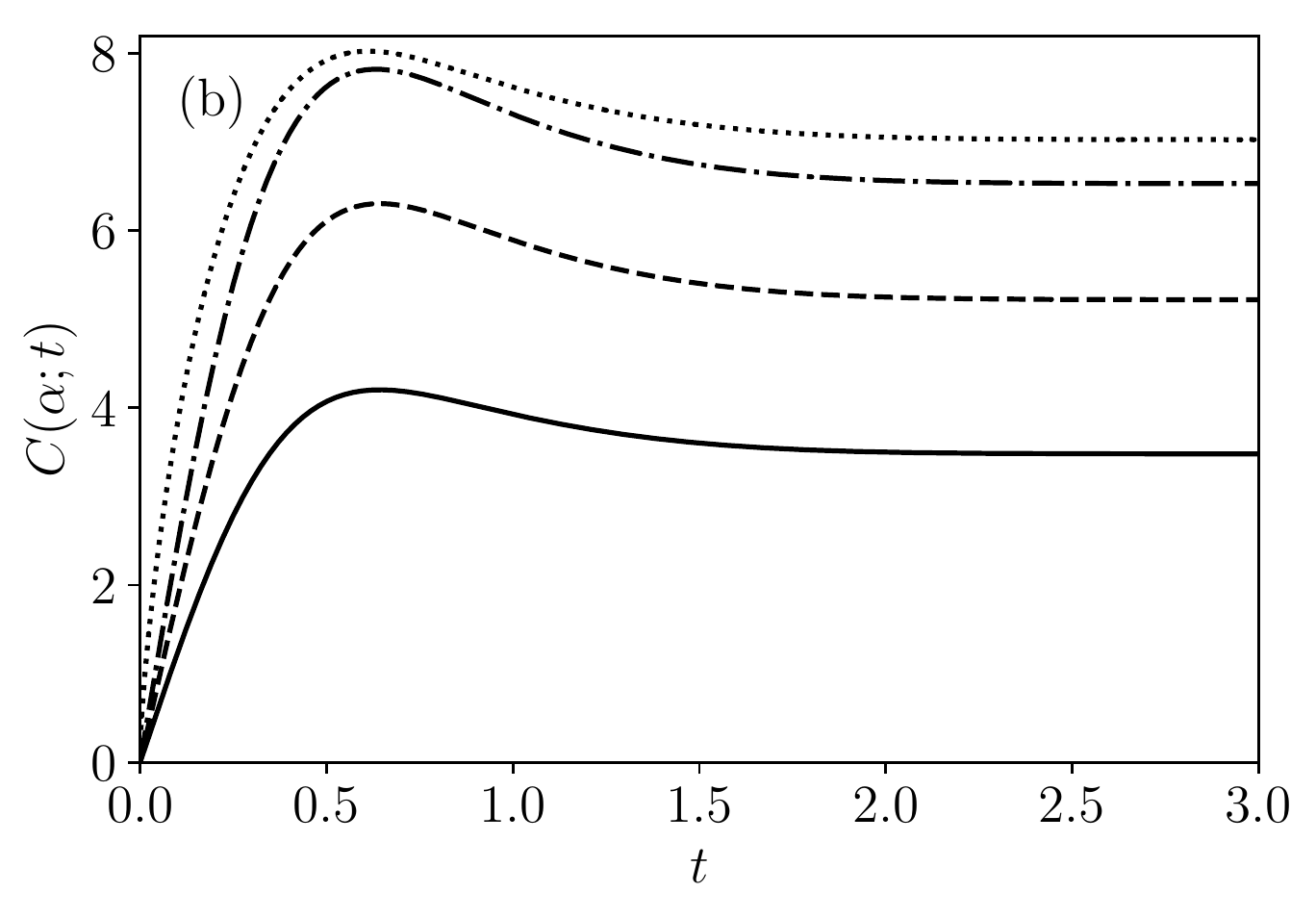}

\caption{Plot of $C(\alpha;t)$ for $\alpha=2$ (solid),
 $\alpha=3$ (dashed), $\alpha=4$ (dot-dashed),
and $\pdv*{C(\alpha;t)}{\alpha}|_{\alpha \rightarrow 1}$ (dotted)
for (a) COE and (b) CUE.}
\label{fig:Calpha}
\end{figure}

Consider the ensemble average of the general moments $\mu_\alpha$, Eq.~\eqref{eq:GenericMoments}, of the Schmidt eigenvalues.  The largest eigenvalue must be separated out from the others and two integrals considered.  First, consider general moments of the sum of all the Schmidt eigenvalues other than the largest, i.e.
\begin{equation}
\overline{ \sum_{l \neq 1} \lambda_l^\alpha(t;\Lambda) } = C_2(\alpha;t) \, \sqrt{\Lambda} \,,
\end{equation}
where after rescaling $s$ to $z$ in Eq.~(\ref{eq:R-function})
\begin{align}
& C_2(\alpha;t) = \int_{-\infty}^\infty \dd{z} \int_0^\infty \dd{w} \overline{R(z,w)} \frac{4^\alpha w^\alpha}{(z^2+4w)^\alpha}  \nonumber \\
& \qquad \qquad \quad \times \sin^{2\alpha}\bigg( \frac{t}{2} \sqrt{z^2+4w}\,\bigg) \nonumber \\
& \qquad \qquad = \int_{-\infty}^\infty \dd{z} \int_0^\infty \dd{w} \rho(w) \frac{4^\alpha w^\alpha}{(z^2+4w)^\alpha}  \nonumber \\
& \qquad \qquad \quad \times \sin^{2\alpha}\bigg( \frac{t}{2} \sqrt{z^2+4w}\,\bigg). \label{eq:C2Integral}
\end{align}
The evaluation of this integral is discussed in the next subsection and
App.~\ref{app:C-2-alpha-t-derivation}.
Now focusing on the ensemble average of the largest Schmidt eigenvalue,
\begin{align}
& \overline{\lambda_1^\alpha(t;\Lambda)} = \overline{\bigg( 1 - \sum_{l \neq 1} \lambda_l(t;\Lambda)\bigg)^\alpha} \nonumber \\
& \qquad\,\,\quad= \overline{\Bigg[ 1 - \int_{-\infty}^\infty \dd{z} \int_0^\infty \dd{w} R(z,w) \frac{4 w}{z^2+4w}} \nonumber \\
& \qquad \qquad \quad \overline{\times \sin^{2\alpha}\bigg( \frac{t}{2} \sqrt{z^2+4w}\,\bigg) \Bigg]^\alpha}\,. \label{eq:LargestEigenvalueExactIntegral}
\end{align}
Before computing the above for a general power $\alpha$, consider the $\alpha =2$ case.  Expanding the above expression gives a square of the integral term, which is a quadruple integral containing the product $R(z_1,w_1) R(z_2,w_2)$.  Equation (\ref{threepoint}) has two contributions, the diagonal term, where $(z_1,w_1) = (z_2,w_2)$, and the off-diagonal one.   For an uncorrelated spectrum, any multi-point spectral correlation function is unity.  Thus
\begin{align}
\overline{\lambda_1^2(t;\Lambda)} = &\; 1 - 2 C_2(1;t)\sqrt{\Lambda} + C_2(2;t) \sqrt{\Lambda} \nonumber \\
&  + C^2_2(1;t)\, \Lambda,
\end{align}
where the off-diagonal term, $R_3(z_1,z_2)$, is responsible for the $\mathcal{O}(\Lambda)$ term.
This illustrates that to the leading $\mathcal{O}(\sqrt{\Lambda})$, the diagonal term alone suffices, and the other terms contribute to higher than leading order.  This simplifies the $\overline{\lambda_1^\alpha}$ computation, where after the binomial expansion of Eq.~(\ref{eq:LargestEigenvalueExactIntegral}), keeping only the terms contributing to the leading order gives,
\begin{align}
& \overline{ \lambda_1^\alpha(t;\Lambda) } = 1 + \sum_{p=1}^\infty (-1)^p \binom{\alpha}{p} \overline{ \sum_{l\neq 1} \lambda_l^p }.
\end{align}
Finally, the general moments Eq.~\eqref{eq:GenericMoments}
of the Schmidt eigenvalues for $\alpha >1/2$ are given by
\begin{equation}
\overline{\mu_{\alpha}(t;\Lambda) } = 1 + \sum_{p=1}^\infty (-1)^p \binom{\alpha}{p} \overline{ \sum_{l\neq 1} \lambda_l^p }\, +\, \overline{ \sum_{l\neq 1} \lambda_l^\alpha } .
\end{equation}
These can be written as
\begin{equation} \label{eq:mu-alpha-t---C-alpha-t}
\overline{ \mu_{\alpha}(t;\Lambda) } = 1 - C(\alpha;t) \sqrt{\Lambda},
\end{equation}
where
\begin{equation} \label{eq:C-alpha-t-via-C-2}
C(\alpha;t) = \sum_{p=1}^\infty (-1)^{p+1} \binom{\alpha}{p} C_{2}(p;t) - C_{2}(\alpha;t).
\end{equation}
These functions are central for the analytical
description of the entropies and shown
in Fig.~\ref{fig:Calpha} for the COE and the CUE.
A particular feature is the overshooting before the saturation sets in.
For the CUE the location of the maxima occurs slightly later in $t$
than for the COE and also the saturation regime is reached
slightly later.
Moreover, the saturation value is slightly larger than in the COE case.

In addition, it can be shown that $\overline{ \mu_\alpha (t;\Lambda) }$ is evaluated up to and including $\mathcal{O}(\sqrt{\Lambda})$ by the first and second largest Schmidt eigenvalues
\begin{equation}
  \overline{ \mu_\alpha (t;\Lambda) }
   = \overline{\lambda_1^\alpha(t;\Lambda)} +
     \overline{\lambda_2^\alpha(t;\Lambda)},
\end{equation}
as other Schmidt eigenvalues ($l > 2$) do not contribute to $ \mathcal{O} ( \sqrt{ \Lambda } ) $.
This relation is vital for recursively invoking perturbation theory
in Sec.~\ref{sec:NonPerturbative}
in order to extend the results beyond the perturbative regime.

\subsection{Entropies}

The HCT entropies can be computed in the perturbation regime using the results for the average eigenvalue moments. For $\alpha \neq 1$ one has
\begin{equation}
\overline{ S_\alpha(t;\Lambda)} = \frac{C(\alpha;t)}{\alpha -1} \sqrt{\Lambda},
\end{equation}
where $C(\alpha; t)$ is given by Eq.~\eqref{eq:C-alpha-t-via-C-2}.
This requires the computation of $C_2(\alpha; t)$,
which is done in App.~\ref{app:C-2-alpha-t-derivation},
and leads to
\begin{equation} \label{eq:C-2-alpha-t}
  C_{2}(\alpha;t) = 2^\alpha \sum_{q=0}^\infty \sum_{m=0}^q (-1)^q
                   \binom{\alpha}{q} a_{qm} f_{m}(\alpha;t),
\end{equation}
where
\begin{equation}
a_{qm} = \binom{q}{\frac{q-m}{2}}
        \left[ \frac{1+(-1)^{q-m}}{2^q(1+\delta_{m,0})} \right]
\end{equation}
and
\begin{align}
  f_m(\alpha;t) = & \int_{-\infty}^\infty \dd{z} \int_0^\infty
              \dd{w} \frac{w^\alpha \rho(w)}{(z^2+4w)^\alpha} \nonumber \\
                  & \qquad \times \cos(m t \sqrt{z^2+4w}\,).
\end{align}
Explicit expressions for $f_m(\alpha;t)$
for the COE and CUE are derived in App.~\ref{app:C-2-alpha-t-derivation-COE}
and \ref{app:C-2-alpha-t-derivation-CUE}, respectively.

\subsection{Discussion}

To discuss some qualitative features, the case $\alpha = 2$, which corresponds
to the linear entropy, is considered here.
By Eqs.~(\ref{eq:GenericHCTEntropy}, \ref{eq:mu-alpha-t---C-alpha-t})
one has $S_2(t;\Lambda) = C(2; t) \sqrt{\Lambda}$.
In case of the COE
\begin{align}
& C(2;t) = 4 \pi t \big( \ue^{-t^2} [ \{ 1 + 2 t^2 \} I_0(t^2) + 2 t^2 I_1(t^2) ] \nonumber \\
& \qquad \qquad - 4 t^2 \ue^{-4 t^2} [ I_0(4 t^2) + I_1(4 t^2) ] \, \big),
\end{align}
where $I_n(z)$ is the modified Bessel function of the first kind
\cite[Eq.~10.25.2]{DLMF}.
Whereas for the CUE case
\begin{align}
& C(2;t) = \pi t \Big( 3 \ue^{-t^2} - \frac{1}{2}\ue^{-4 t^2} \Big) + \pi^{3/2} \text{erf}(t) \Big( \frac{1}{2} + 3 t^2 \Big) \nonumber \\
& \qquad \qquad + \pi^{3/2} \text{erf}(2 t) \Big( \frac{1}{8} - 3 t^2 \Big),
\end{align}
where $\text{erf}(z)$ is the error function.
For both COE and CUE cases, $C(2;t)$ for small $t$ has the expansion
\begin{align} \label{eq:C-2-t}
  & C(2;t) = 4 \pi t + \mathcal{O}(t^3)
\end{align}
and naturally gives linear-in-time entropy growth for short time $t$.
The same is true for other $\alpha$-entropies, except for $\alpha = 1$,
for which the leading term is of the order $\mathcal{O}(t\,\ln t)$. In fact, it can be shown that for both COE and CUE cases, with $\alpha > 1$ and short time $t$,
\begin{equation}
\dv{}{t} C(\alpha;t) \big|_{t \rightarrow 0} = 2 \pi \alpha. \label{eq:InitialRateCalpha}
\end{equation}

In the limit $t \rightarrow \infty$, saturation values of the entropies
can be obtained from
\begin{equation}
   S_2(\infty; \Lambda) = \frac{C(\alpha; \infty)}{\alpha-1} \sqrt{\Lambda},
\end{equation}
which are of the order $\mathcal{O}(\sqrt{\Lambda}\,)$.
Using the explicit expressions for $f_m(\alpha;t)$
derived in App.~\ref{app:C-2-alpha-t-derivation-COE}, \ref{app:C-2-alpha-t-derivation-CUE}
one sees that in the limit $t \rightarrow \infty$,
$f_m(\alpha;t)$ vanish for all $m \neq 0$.
 Using this fact, an expression for saturation value
of the $\alpha$-entropies ($\alpha \neq 1$) can be derived as
\begin{align}
\overline{S_\alpha(\infty, \Lambda)} & = \Bigg(\alpha \,\,  _{3}F_2(1/2,3/2,1-\alpha\,;\,2,2\,;\,1) \nonumber \\
&\; \qquad \quad - \frac{2}{\pi} \,\,\frac{\Gamma(\alpha-1/2)\,\Gamma(\alpha+1/2)}{\Gamma(\alpha)\,\Gamma(\alpha+1)} \Bigg) \nonumber \\
& \; \qquad \times \frac{\sqrt{\Lambda}}{\alpha-1}\begin{cases}
							\sqrt{2\pi} \; \quad \quad  \text{for COE,} \\
							\pi^{3/2}/2 \, \; \quad \text{for CUE.}
						 \end{cases}
  \label{eq:sat-S-alpha}
\end{align}
Here $_{m}F_n$ is a generalized hypergeometric function
\cite[Eq.~35.8.1]{DLMF} defined by
\begin{align}
& _{m}F_n ( a_1 ,\ldots, a_m ; b_1 ,\ldots, b_n ; z) = \sum_{k=0}^\infty \frac{(a_1)_k \ldots (a_m)_k}{(b_1)_k \ldots (b_n)_k} \frac{z^k}{k!},
\end{align}
where $(a)_k = \Gamma(a+k)/\Gamma(a)$ is Pochhammer's symbol.
Equation~\eqref{eq:sat-S-alpha} shows that the saturation values
for both COE and CUE scale with $\sqrt{\Lambda}$
and that the CUE case leads to a slightly (11\%) larger value.

For the linear entropy, $\alpha=2$, Eq.~\eqref{eq:sat-S-alpha} simplifies to
\begin{align} \label{eq:S2-sat-COE-CUE}
\overline{S_2(\infty;\Lambda)} = \sqrt{\Lambda}
  \begin{cases}
  5 \sqrt{\pi/8}   & \text{for COE}, \\
  5 \pi^{3/2}/8   & \text{for CUE}.
  \end{cases}
\end{align}
For the special case of the von Neumann entropy for $\alpha=1$,
$\lim_{t\rightarrow \infty} \pdv*{C(\alpha;t)}{\alpha}|_{\alpha \rightarrow 1}$
needs to be computed. It can be shown that
\begin{align} \label{eq:saturation-perturbatively}
\overline{S_1(\infty;\Lambda)}
= & \Big(
 4 \ln 2- \frac{3}{16}\, _{4}F_3(1,1,3/2,5/2 \,;\,2,3,3\,;\,1)   \Big)
\nonumber \\
& \times \sqrt{\Lambda} \begin{cases}
							\sqrt{2\pi} \; \quad \quad  \text{for COE,} \\
							\pi^{3/2}/2 \, \; \quad \text{for CUE.}
						 \end{cases}
\end{align}
An extension to the non-perturbative result will be discussed
in Sec.~\ref{sec:long-time-saturation}.

\begin{figure}
    \centering
    \includegraphics[width=8.6cm]{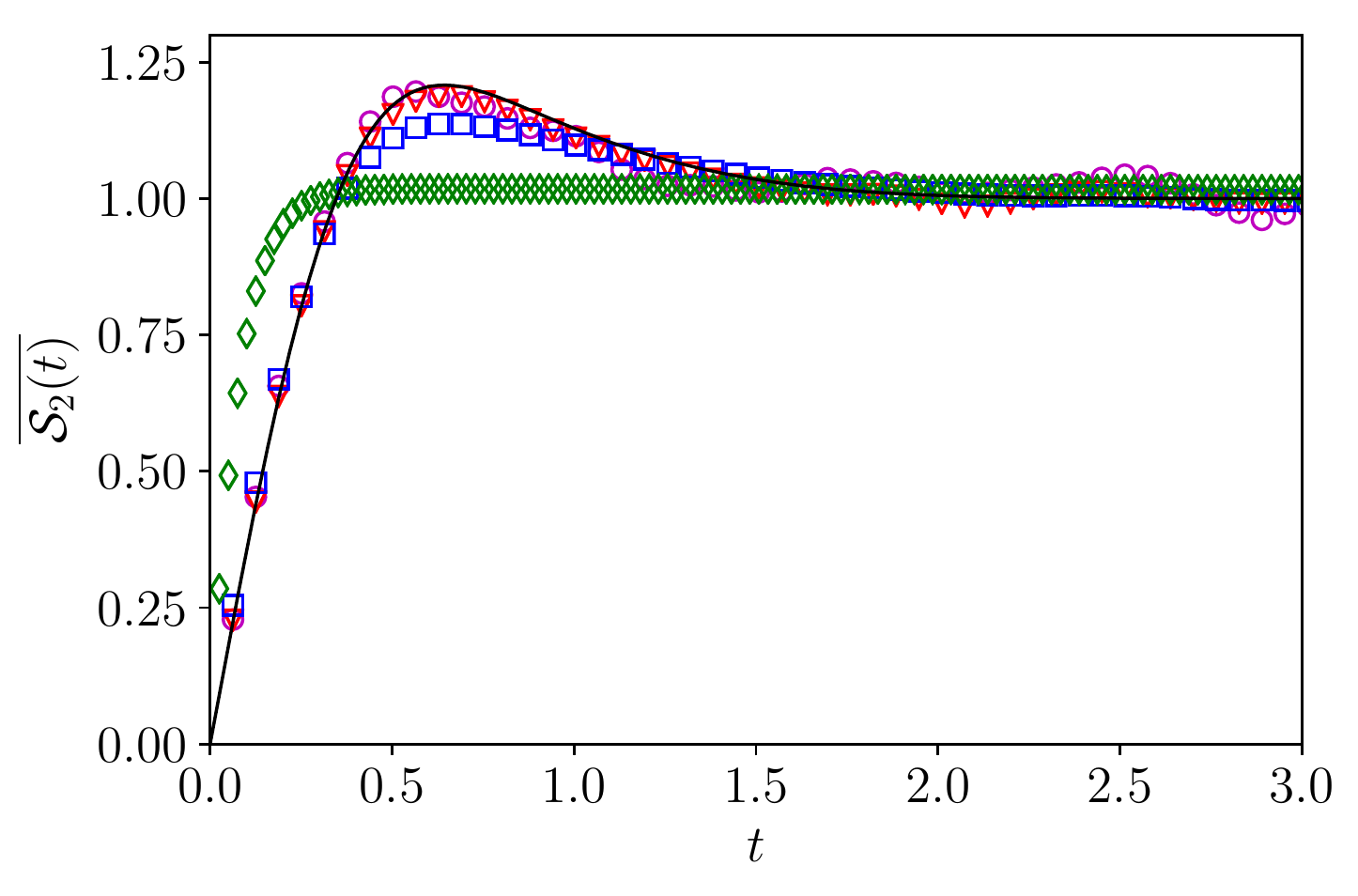}

    \caption{\label{fig:UniversalCurvePlot} Scaled linear entropy
      $\overline{ \mathcal{S}_2(t) }$, Eq.~\eqref{eq:Universal-Scaling}, for the
      random matrix transition ensemble in Eq.~(\ref{eq:GenericFloquetRMT}) for
      the CUE case for $\Lambda = 10^{-6}$ (magenta circles),
      $\Lambda = 10^{-4}$ (red triangles), $\Lambda = 10^{-2}$ (blue squares),
      and $\Lambda = 1$ (green diamonds); The theoretical prediction
      Eq.~(\ref{eq:UniversalCurveTheory}) for $\alpha =2$ is shown as solid curve.}
\end{figure}

In the perturbative regime, if the entropies $\overline{ S_\alpha (t;\Lambda)}$ are scaled with respect to
their saturation values,
\begin{equation}
\overline{ \mathcal{S}_\alpha (t) }
    = \frac{\overline{S_\alpha(t; \Lambda)}}
           {\overline{S_\alpha(\infty; \Lambda)}},
\label{eq:Universal-Scaling}
\end{equation}
they do not depend on the transition parameter,
leading to one universal curve for each $\alpha$ described
by the prediction
\begin{equation}
\overline{ \mathcal{S}_\alpha (t) } = \frac{C(\alpha;t)}{C(\alpha;\infty)}.
  \label{eq:UniversalCurveTheory}
\end{equation}
This universal property is depicted for the linear entropy in Fig.~\ref{fig:UniversalCurvePlot} for various $\Lambda$-values. As $\Lambda$ goes beyond the perturbation regime, departure from the universal curve is seen due to the breakdown of the perturbation theory. In the forthcoming section, the extension of the theory to the non-perturbative regime is discussed.

\section{Non-perturbative regime} \label{sec:NonPerturbative}

The results obtained from the perturbation theory can be extended to the
non-perturbative regime to produce the full transition and the saturation
values by employing the recursively embedded perturbation theory technique as
done in Ref.~\cite{Lakshminarayan16,Tomsovic18c} for the eigenstates.

\subsection{Full transition}

For small enough $\Lambda$, the time evolved state $\ket{jk(n;\epsilon)} \equiv \ket{jk(t;\Lambda)}$ can be Schmidt decomposed as
\begin{equation}
\ket{jk(t;\Lambda)} = \sqrt{\lambda_1(t;\Lambda)} \,\ket{(jk)_1} + \sqrt{\lambda_2(t;\Lambda)}  \,\ket{(jk)_2},
\end{equation}
such that $\lambda_1 + \lambda_2 = 1$, where the time-dependent phase-factor is absorbed into the definition of the Schmidt eigenvectors $\ket{ (jk)_l }$. Now increasing the interaction strength, another unperturbed state energetically close to $\ket{(jk)_1}$ will contribute to $\ket{jk(n;\epsilon)}$,
\begin{align}
& \ket{jk(t;\Lambda)} = \sqrt{\lambda_1'(t;\Lambda)} \,\bigg(\sqrt{\lambda_1(t;\Lambda)} \,\ket{(jk)_1} + \nonumber \\
& \qquad \qquad \quad \sqrt{\lambda_2(t;\Lambda)}  \,\ket{(jk)_2}\bigg) + \sqrt{\lambda_2'(t;\Lambda)} \, \ket{(jk)_3},
\end{align}
where $\lambda_{1,2}'$ follow same statistical properties as the unprimed ones. Thus the purity is $\mu_2' = \lambda_1^{'2} \lambda_1^2 + \lambda_1^{'2} \lambda_2^2 + \lambda_2^{'2}$ giving
\begin{align}
\mu_2' - \mu_2 = -(1-\lambda_1^{'2} - \lambda_2^{'2})\mu_2 + \lambda_2^{'2} (1-\mu_2).
\end{align}

\begin{figure}
\includegraphics[width=8.4cm]{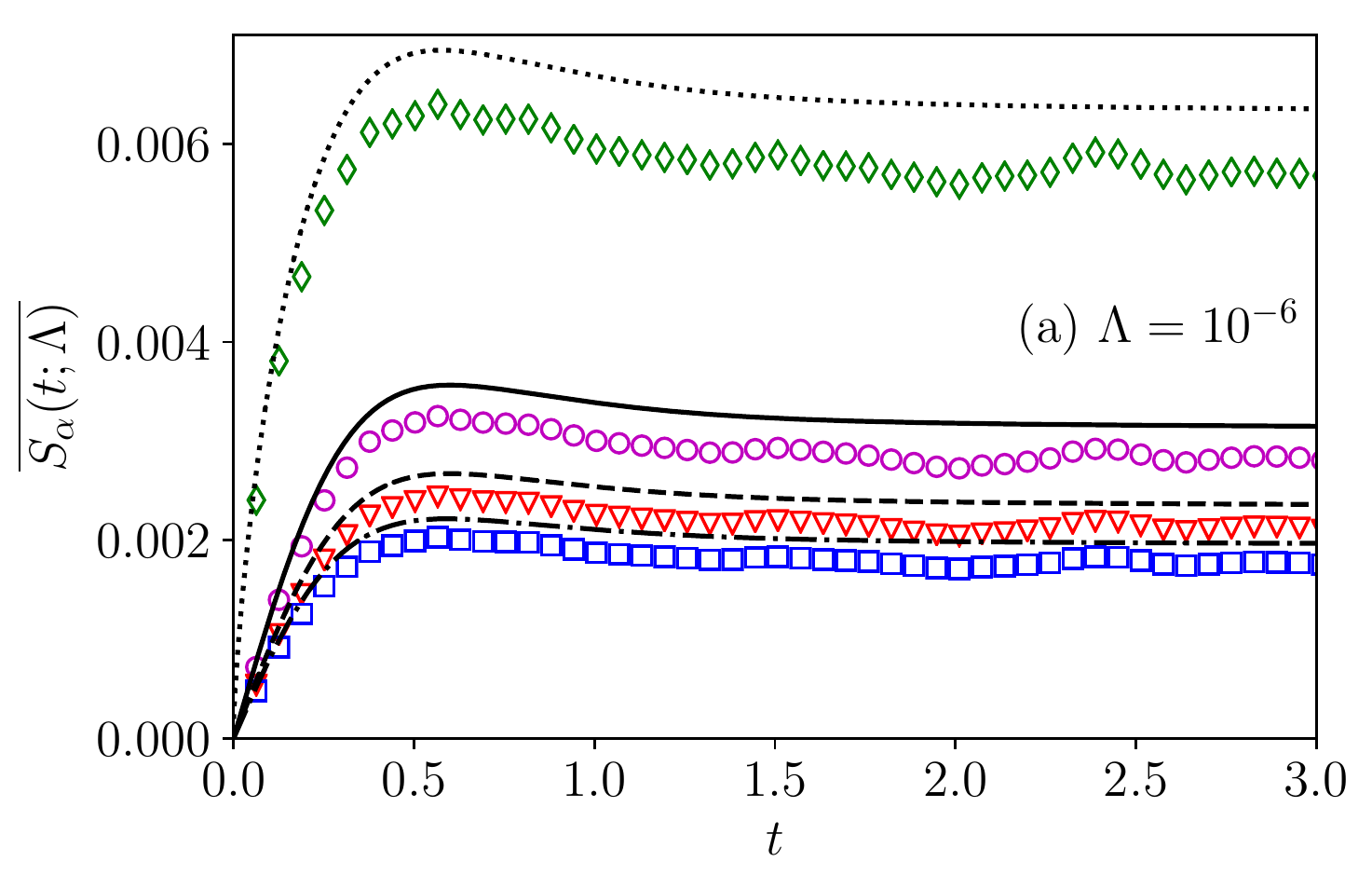}

\includegraphics[width=8.4cm]{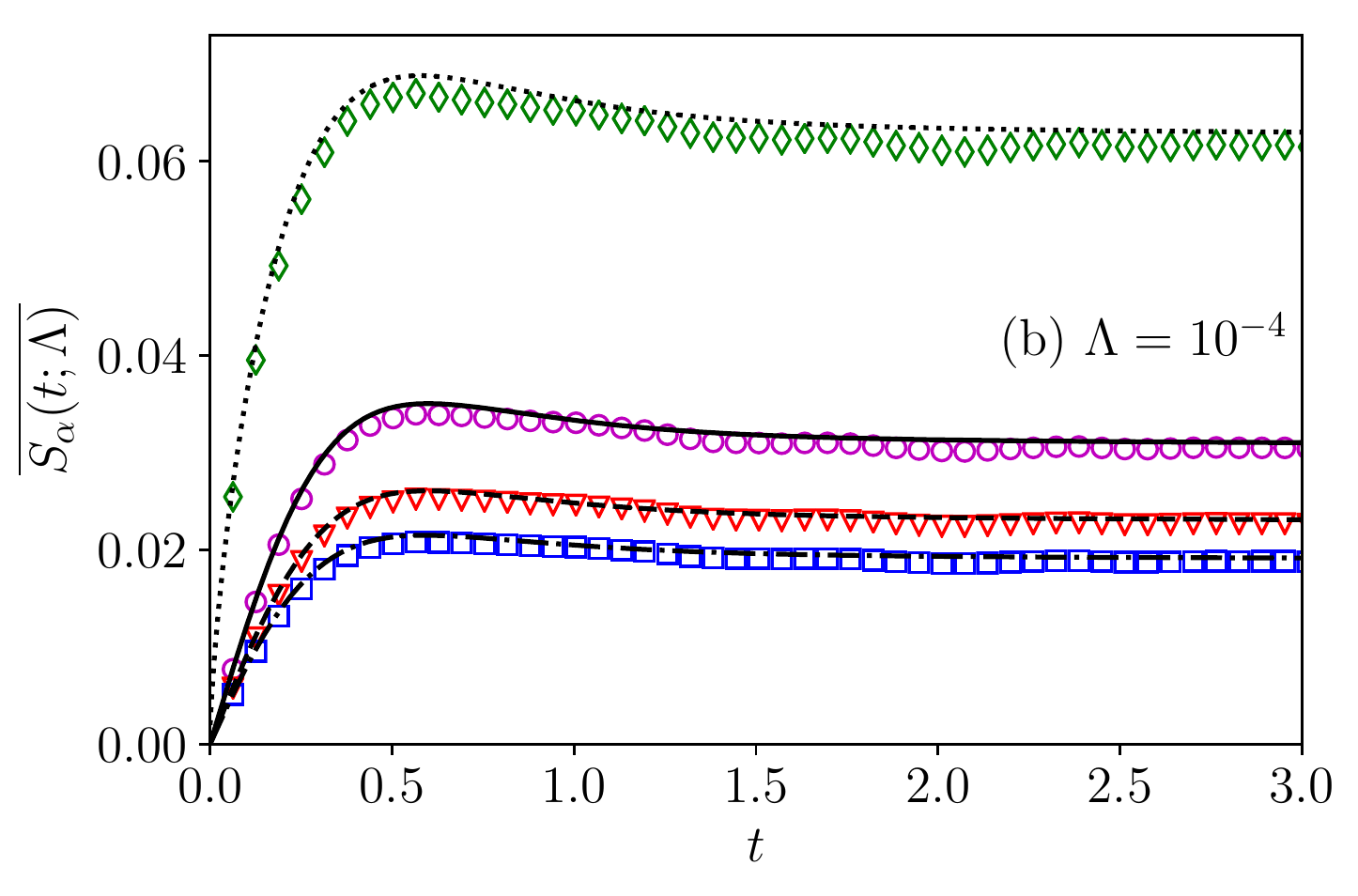}

\includegraphics[width=8.4cm]{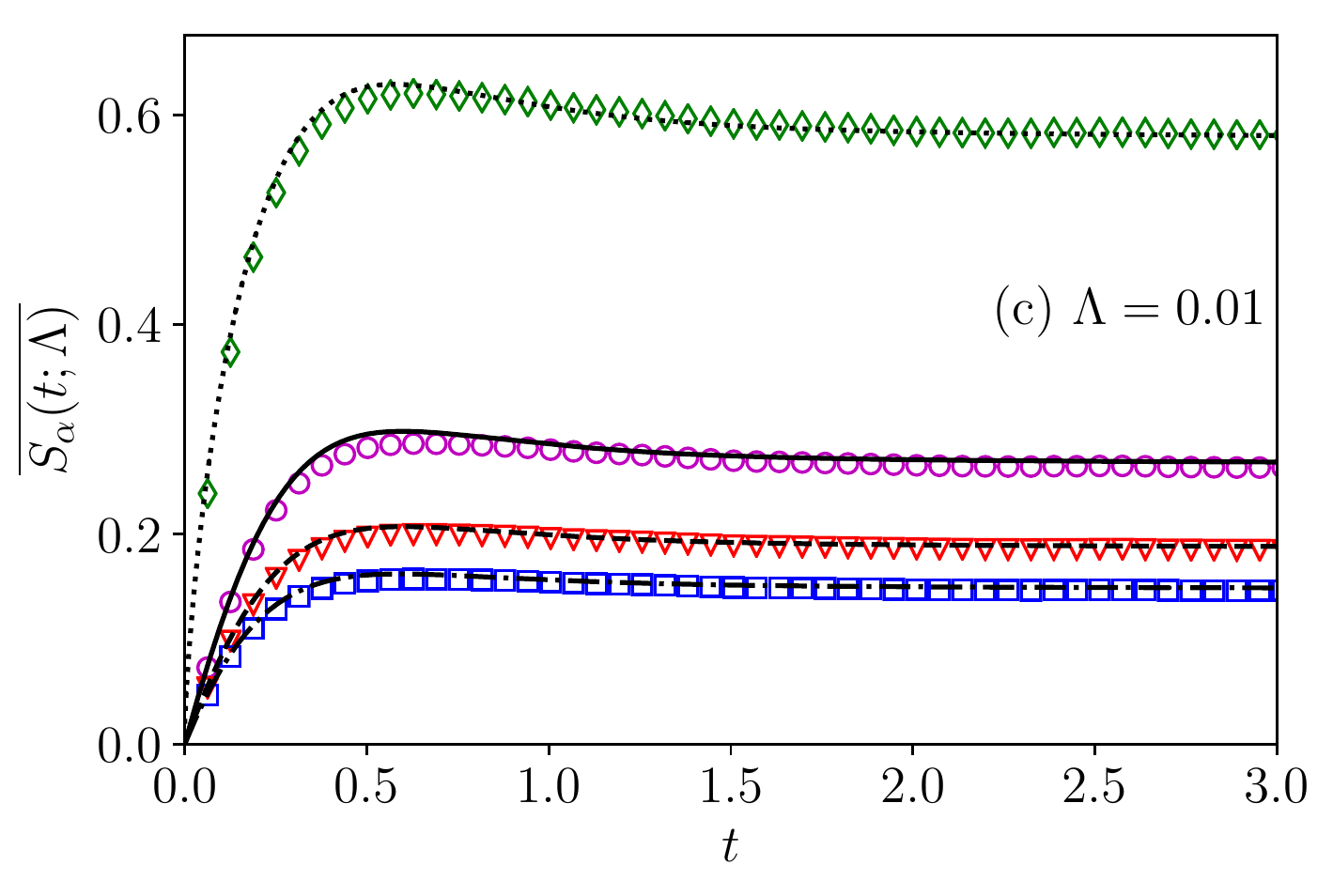}

\includegraphics[width=8.4cm]{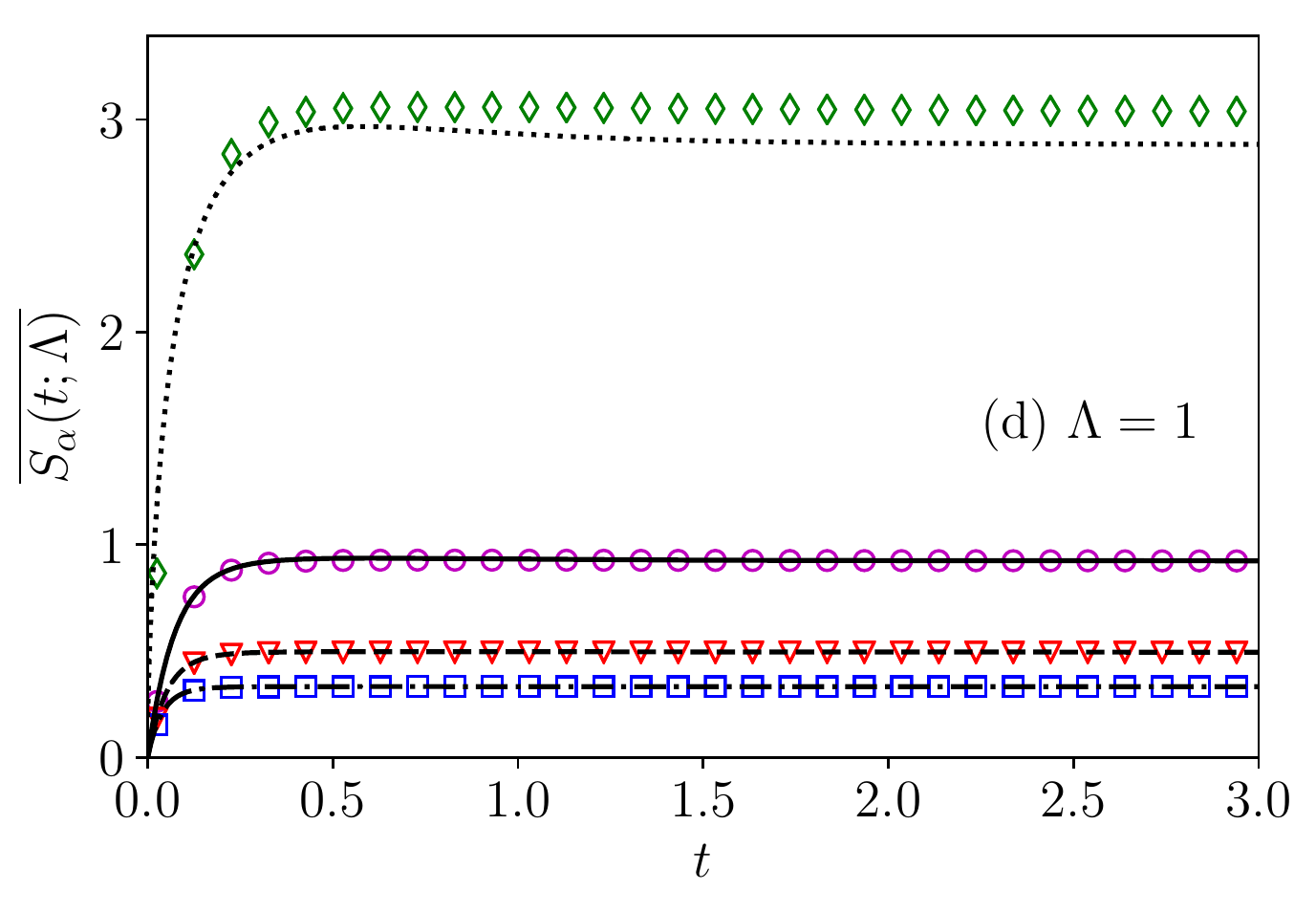}
\caption{\label{fig:COE_Salpha} Entropies $\overline{ S_\alpha }$ for the COE
  case with $N_A=N_B=50$ for (a) $\Lambda=10^{-6}$, (b) $\Lambda=10^{-4}$, (c)
  $\Lambda=10^{-2}$, and (d) $\Lambda=1$ for $\alpha=1$ (green diamonds),
  $\alpha=2$ (magenta circles), $\alpha=3$ (red triangles), and $\alpha=4$
  (blue squares). Black lines show the corresponding theory curves,
  Eq.~(\ref{eq:alphaEntropyTheory}).}
\end{figure}

\begin{figure}
\includegraphics[width=8.4cm]{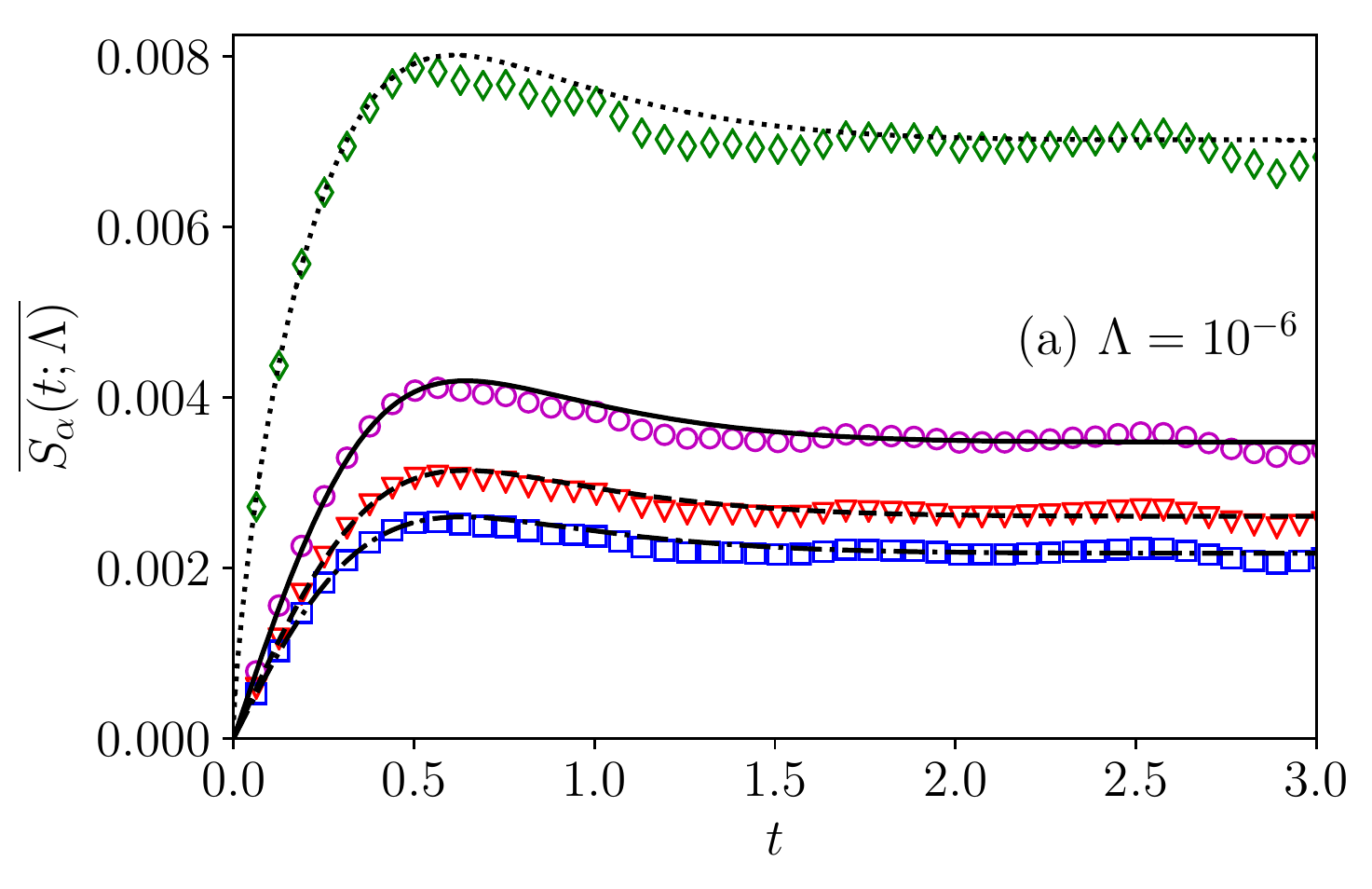}

\includegraphics[width=8.4cm]{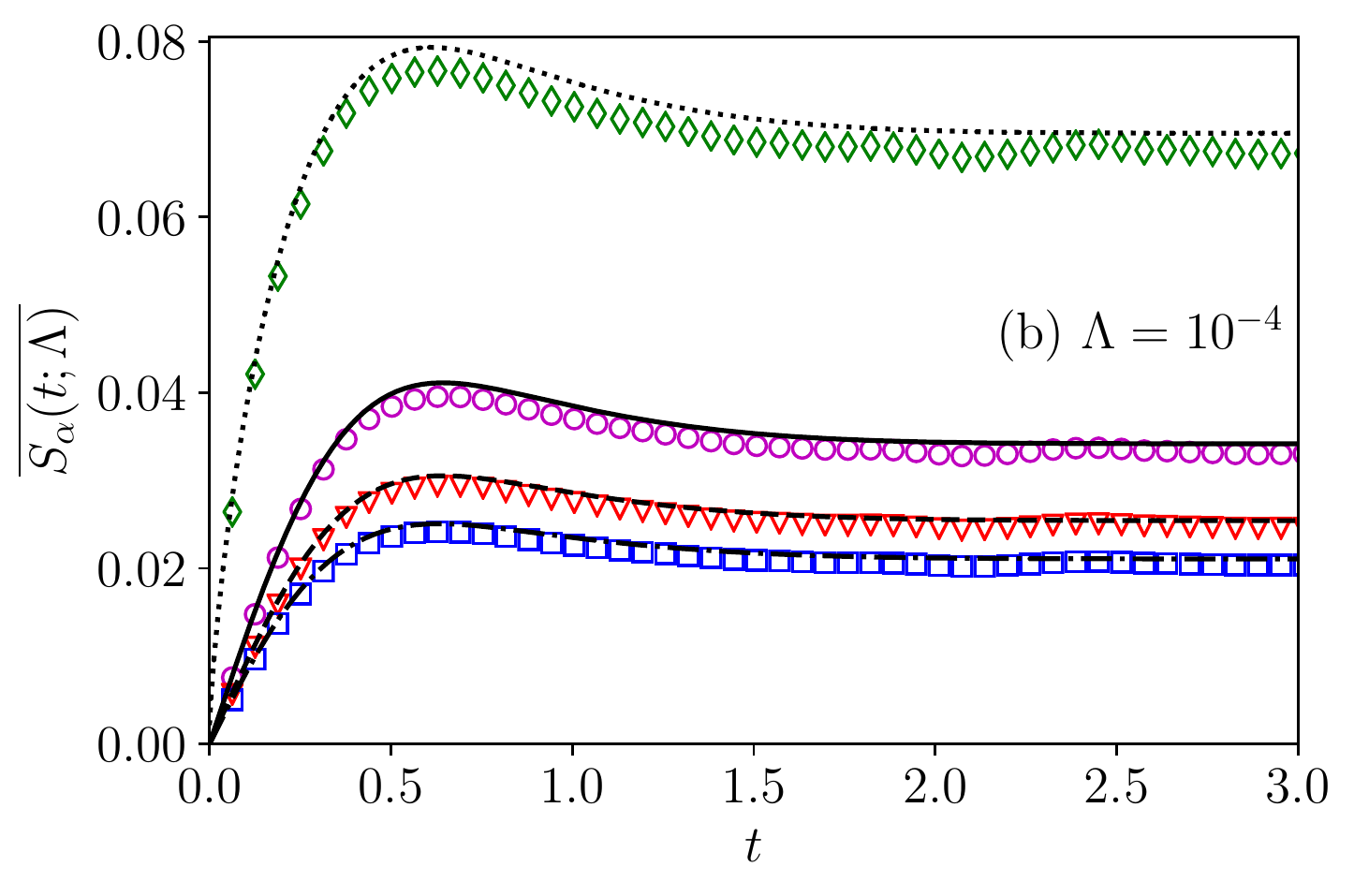}

\includegraphics[width=8.4cm]{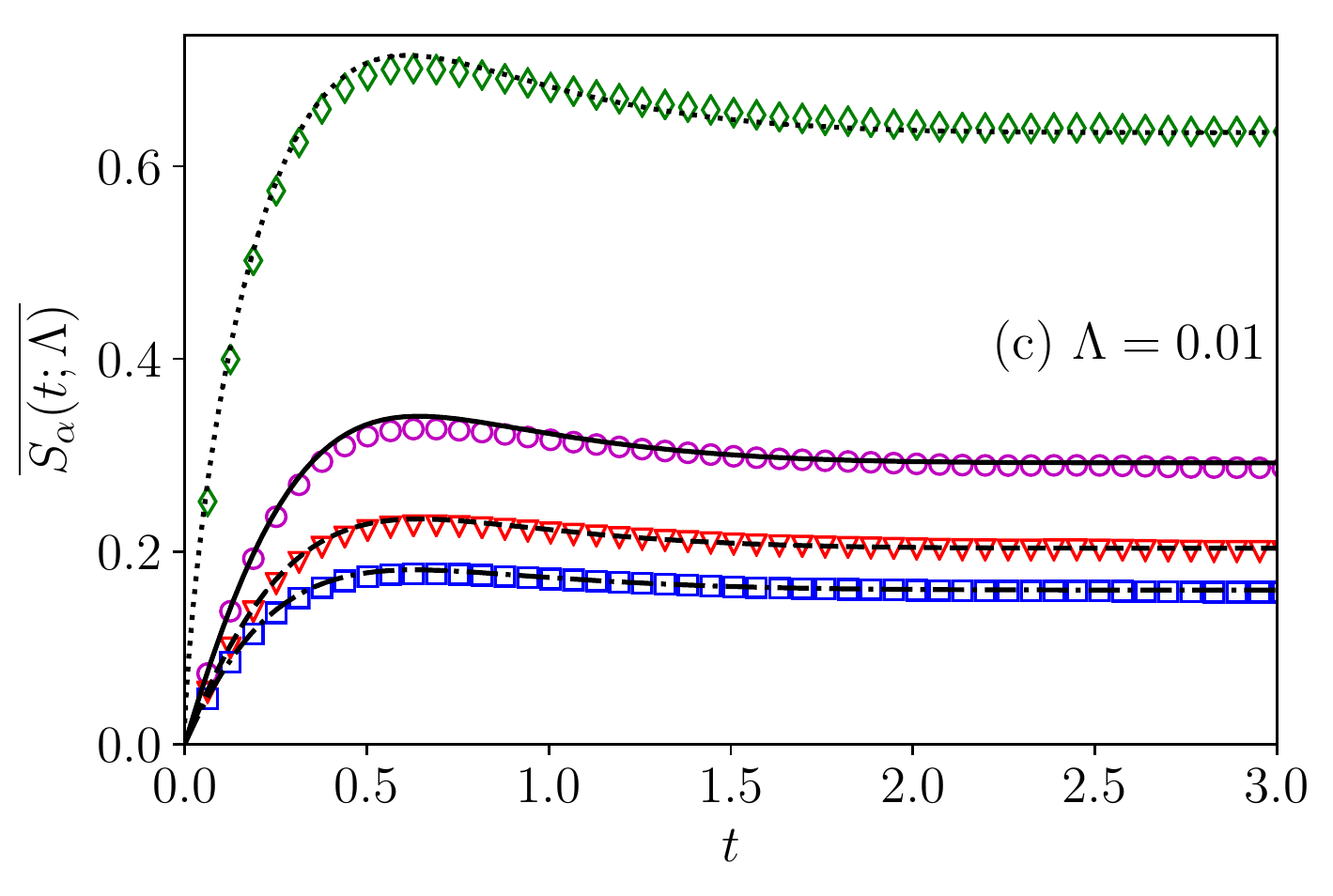}

\includegraphics[width=8.4cm]{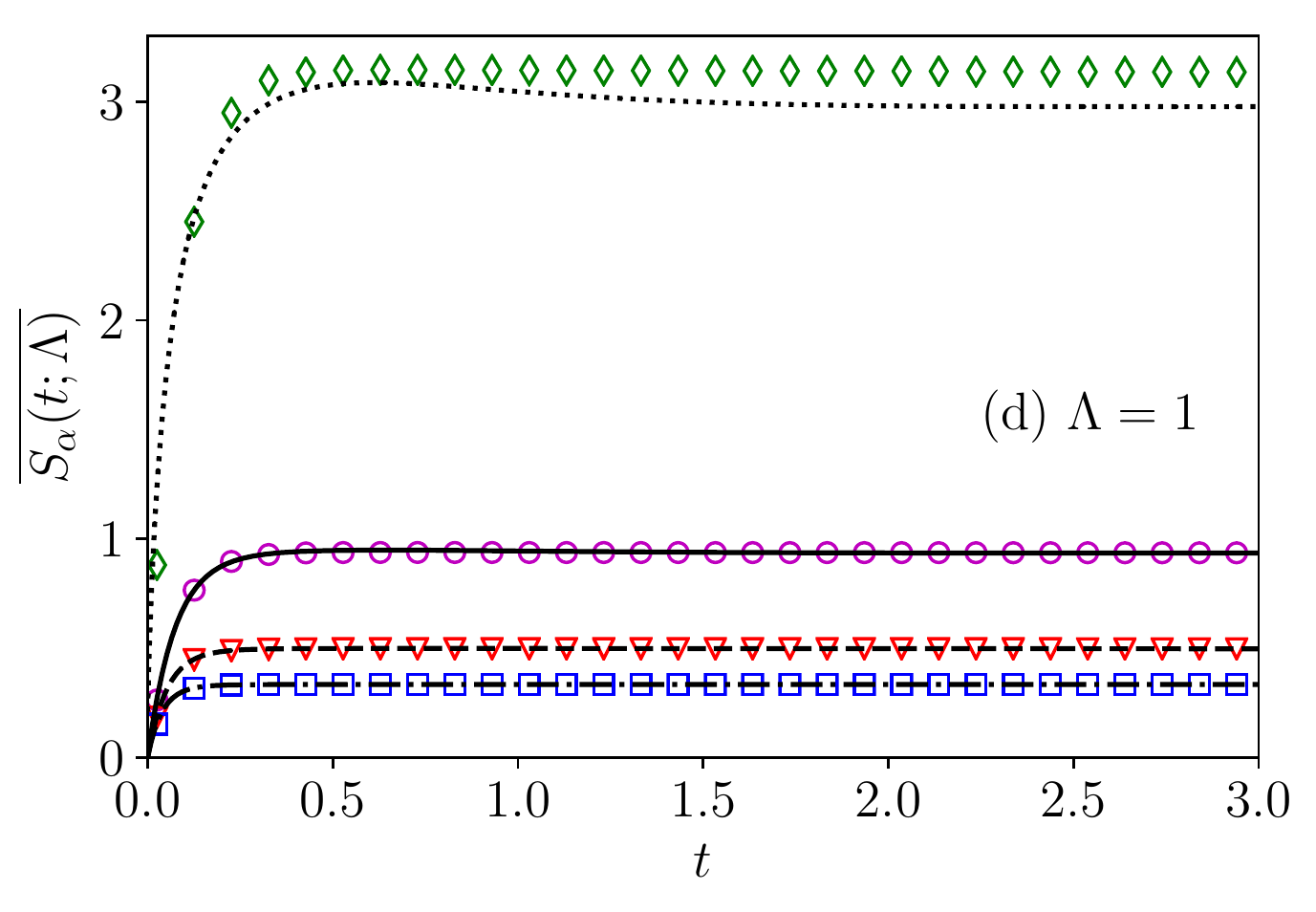}
\caption{\label{fig:CUE_Salpha} Entropies $\overline{ S_\alpha }$ for the CUE
  case with $N_A=N_B=50$ for (a) $\Lambda=10^{-6}$, (b) $\Lambda=10^{-4}$, (c)
  $\Lambda=10^{-2}$, and (d) $\Lambda=1$ for $\alpha=1$ (green diamonds),
  $\alpha=2$ (magenta circles), $\alpha=3$ (red triangles), and $\alpha=4$
  (blue squares). Black lines show the corresponding theory curves,
  Eq.~(\ref{eq:alphaEntropyTheory}).}
\end{figure}

For a given $\alpha$, following this technique and replacing $\lambda_{1,2}^{'\alpha}$ with their ensemble average, a differential equation for the moments $\overline{ \mu_\alpha (t;\Lambda) }$ can be derived, good up to $\mathcal{O}(\sqrt{\Lambda})$,
\begin{equation}
\pdv{\overline{ \mu_\alpha(t;\Lambda) }}{\sqrt{\Lambda}} = - C(\alpha;t) \overline{ \mu_\alpha (t;\Lambda)}.
\end{equation}
This has a solution of the form (valid for the infinite dimensional case)
\begin{equation}
\overline{ \mu_\alpha(t;\Lambda)} \approx \exp(-C(\alpha;t) \sqrt{\Lambda}).
\end{equation}
In the limit $\Lambda \rightarrow \infty$, and for large (but finite)
dimensionality $N=N_A =N_B$, the moments tend to the random matrix result
\begin{equation}
\overline{ \mu_\alpha^\infty } = \mathcal{C}_\alpha/N^{\alpha-1},
\end{equation}
where the $\mathcal{C}_\alpha$ are Catalan numbers \cite[\S26.5.]{DLMF}.
For the $N_A \neq N_B$ case such an expression can be found following Ref.~\cite{Sommers04}. Incorporating this limit
into the above differential equation solution gives an approximate expression for the moments valid for any $\Lambda$,
\begin{equation}
\overline{ \mu_\alpha (t;\Lambda) } \approx \exp(- \frac{C(\alpha;t)}{1-\overline{\mu_\alpha^\infty}}\sqrt{\Lambda})(1-\overline{\mu_\alpha^\infty}) + \overline{ \mu_\alpha^\infty}.
\end{equation}
Using the definition of the HCT entropies (\ref{eq:GenericHCTEntropy})
gives
\begin{equation}
\overline{S_\alpha(t;\Lambda) } \approx \bigg[ 1 - \exp(-\frac{C(\alpha;t)}{(\alpha-1)\overline{ S_\alpha^\infty}}\sqrt{\Lambda})\bigg] \overline{S_\alpha^\infty}, \label{eq:alphaEntropyTheory}
\end{equation}
where
\begin{equation}
\overline{ S_\alpha^\infty} = \frac{1-\mathcal{C}_\alpha N^{1-\alpha}}{\alpha-1}.
\end{equation}
To apply Eq.~\eqref{eq:alphaEntropyTheory}
one has to use for $C(\alpha; t)$ the results
corresponding to the CUE or the COE,
as given by Eq.~\eqref{eq:C-alpha-t-via-C-2}.
When $\Lambda$ is large, however,
there is no difference between CUE and COE
due to the same scaling of $C(\alpha; t)$,
as for example in Eq.~\eqref{eq:C-2-t} for $\alpha=2$.

The result Eq.~(\ref{eq:alphaEntropyTheory}) is in agreement with
numerical computations for both the COE, see Fig.~\ref{fig:COE_Salpha},
and the CUE, see Fig.~\ref{fig:CUE_Salpha}.
For these numerical calculations, 20 realizations of the random matrix model Eq.~\eqref{eq:GenericFloquetRMT} for $N_A=N_B=50$ have been used, leading to a total of $5 \times 10^4$ initially unentangled eigenstates $\ket{jk}$ used for averaging. This amount of averaging is
particularly relevant for small values of $\Lambda$
for which the time evolution of the entanglement of the individual
states shows strong fluctuations from one state to another.
These are also the origin of the small fluctuations
seen in both figures for $\Lambda=10^{-6}$
for the von Neumann entropy $\overline{S_1(t; \Lambda)}$,
which is the most sensitive of the considered entropies.
Moreover, at small $\Lambda$, finite $N$ effects
become visible, in particular for the COE,
due to the small overall amount of entanglement.
Increasing the matrix dimension of the subsystems
to $N=100$ improves the agreement with the theoretical prediction
(not shown).
For $\Lambda=10^{-4}$ and $\Lambda=10^{-2}$ excellent
agreement of the numerically computed entropies and
the theory is found.
For $\Lambda=1$ again the von Neumann entropy shows
small deviations from the theoretical prediction.

\subsection{Long-time saturation}
\label{sec:long-time-saturation}

\begin{figure}[b]
\centering
 \includegraphics[width=8.4cm]{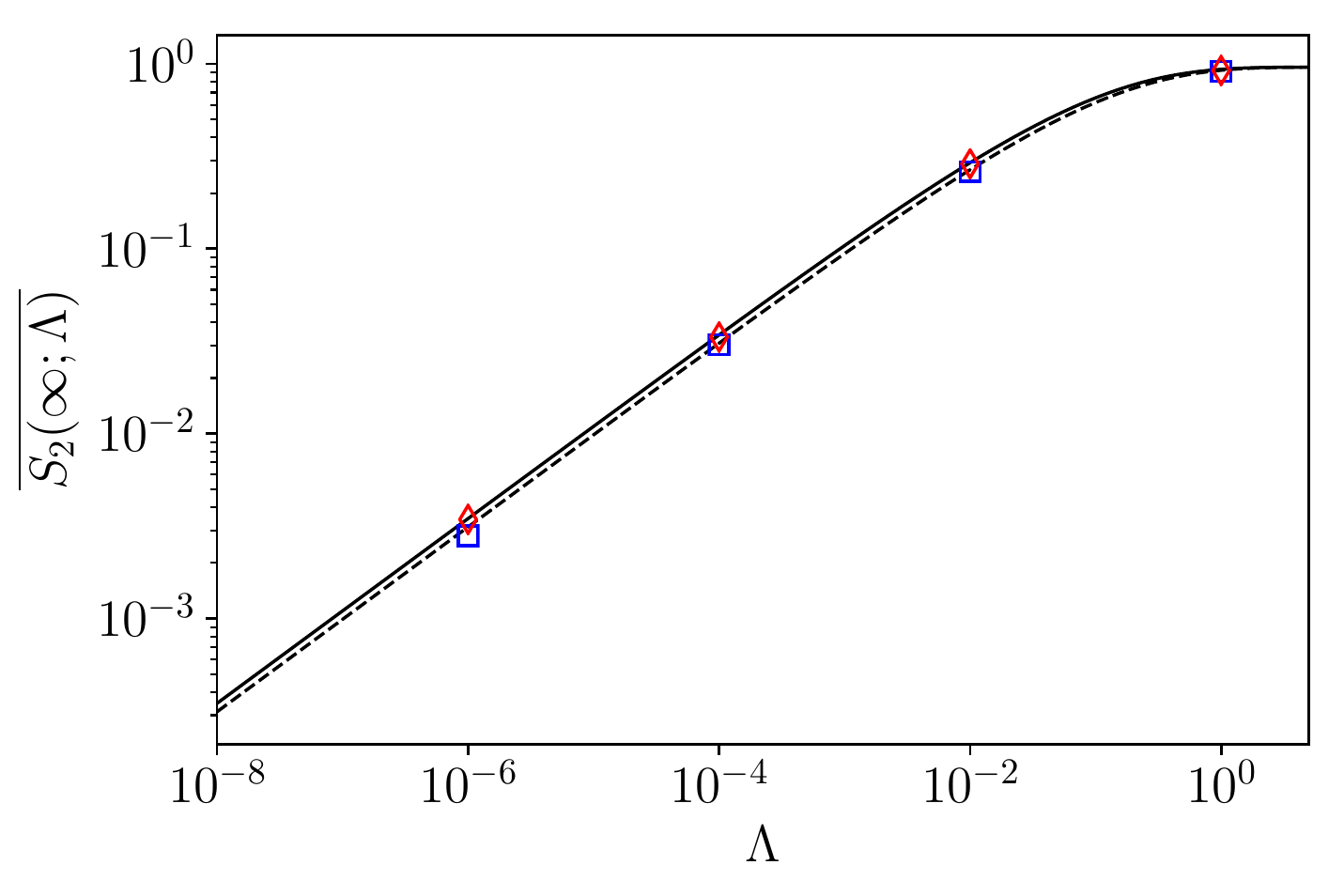}
\caption{\label{fig:saturation}
Saturation values of the linear entropy, $\overline{S_2(\infty;\Lambda)}$, as a function of $\Lambda$
for the COE (blue squares) and CUE (red diamonds) in comparison
with the prediction Eq.~\eqref{eq:sat-alphaEntropyTheory}
(dashed and solid black lines representing COE and CUE, respectively).}
\end{figure}

Using the result Eq.~\eqref{eq:alphaEntropyTheory}
one can also perform the long-time limit to
obtain a prediction for the saturation
values $\overline{S_\alpha(\infty;\Lambda) }$
going beyond the perturbative result Eq.~\eqref{eq:saturation-perturbatively}.
Thus one gets
\begin{equation}
   \overline{S_\alpha(\infty;\Lambda) } = \bigg[ 1 - \exp(-\frac{C(\alpha;\infty)}{(\alpha-1)\overline{ S_\alpha^\infty}}\sqrt{\Lambda})\bigg] \overline{S_\alpha^\infty}. \label{eq:sat-alphaEntropyTheory}
\end{equation}
For large $\Lambda$ the exponential becomes very small
so that the saturation reaches $\overline{S_\alpha^\infty}$.
However, for small $\Lambda$ a reduced saturation value is obtained.
Figure~\ref{fig:saturation} illustrates
this for the linear entropy for the COE and CUE
where Eq.~\eqref{eq:S2-sat-COE-CUE} is used for $C(\alpha, \infty)$.
Very good agreement of the prediction with the numerical results
is found.
Up to $\Lambda=10^{-1}$ the saturation value
follows Eq.~\eqref{eq:S2-sat-COE-CUE}
and then the behavior given by Eq.~\eqref{eq:sat-alphaEntropyTheory}
sets in.
The saturation values for the COE are below
those of the CUE but eventually both approach $\overline{S_2^\infty}$.

\section{Coupled kicked rotors}
\label{sec:coupled-kicked-rotors}

\begin{figure}
\includegraphics[width=8.4cm]{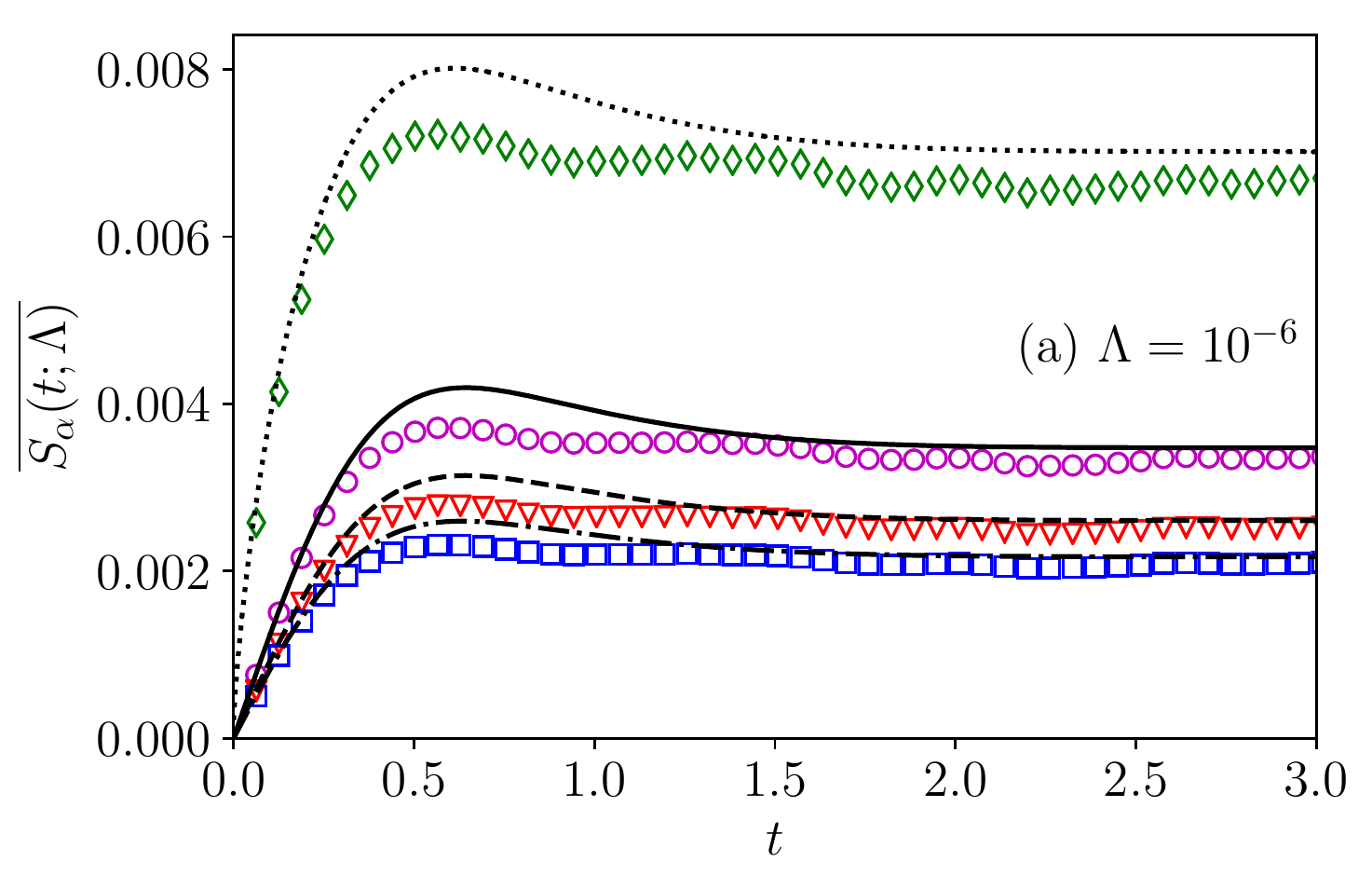}

\includegraphics[width=8.4cm]{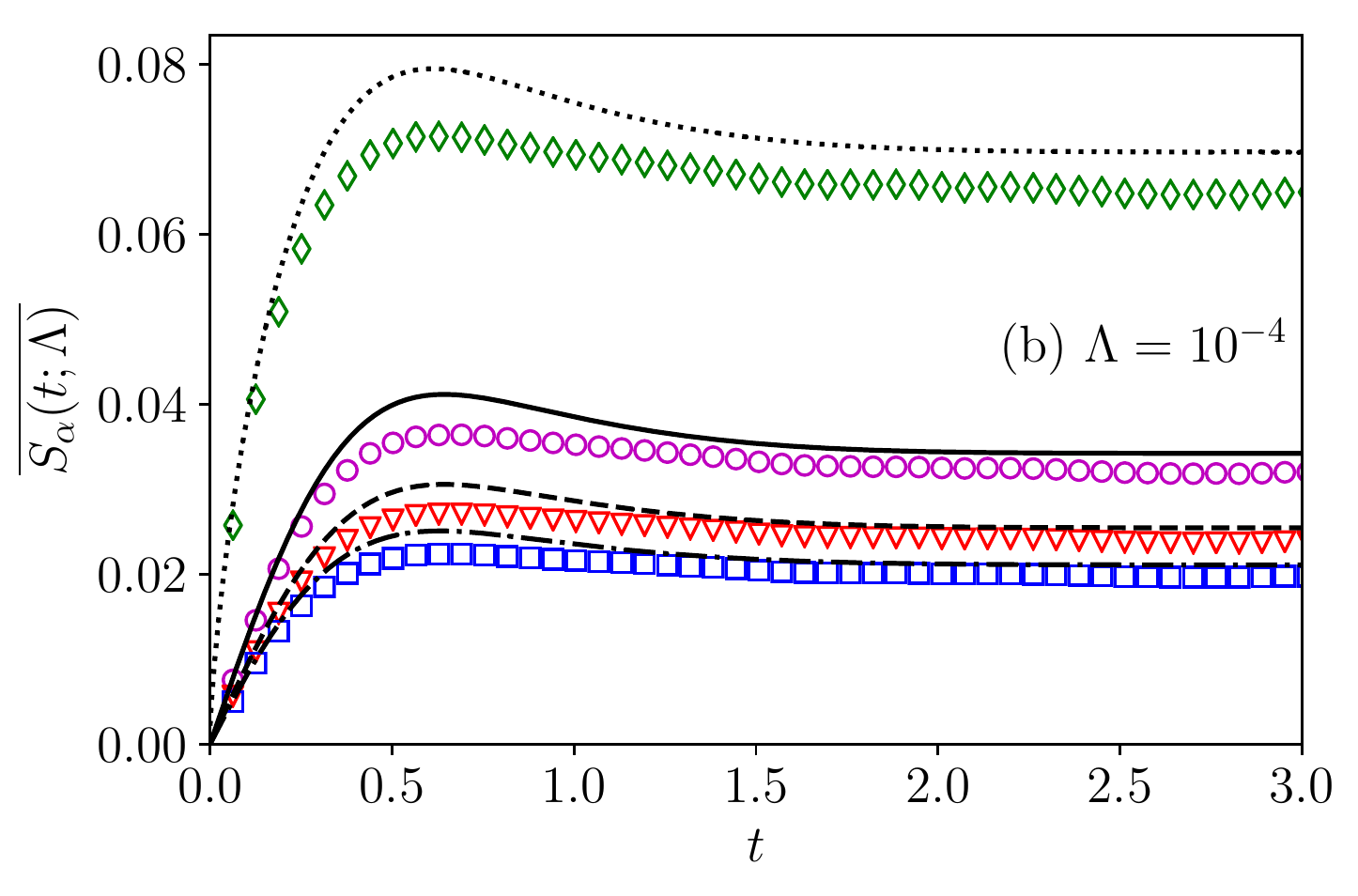}

\includegraphics[width=8.4cm]{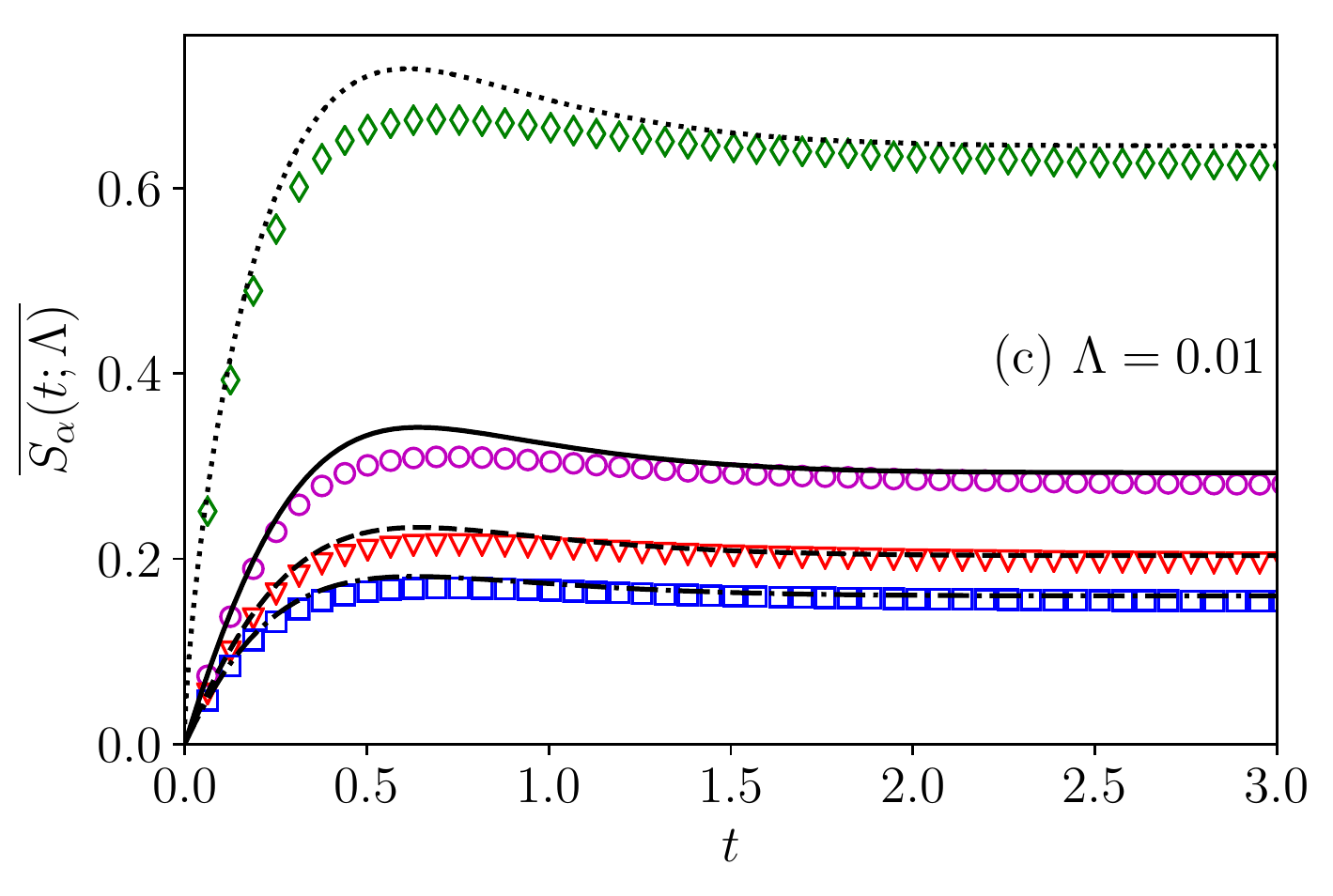}

\includegraphics[width=8.4cm]{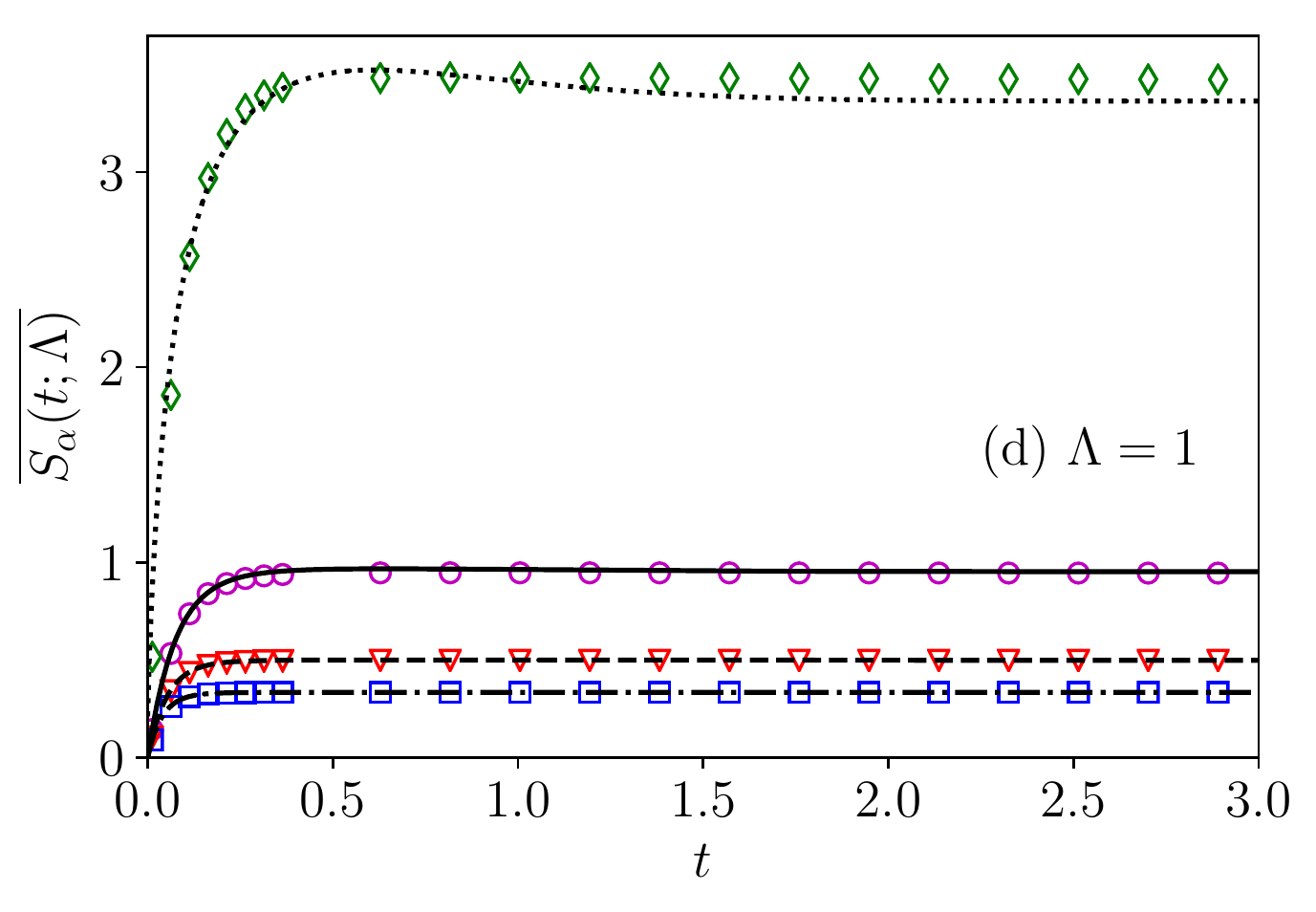}

\caption{\label{fig:KR_Salpha} Entropies $\overline{ S_\alpha }$ for the coupled kicked rotors with completely broken time-reversal invariance and $N=100$, for (a) $\Lambda=10^{-6}$, (b) $\Lambda=10^{-4}$, (c)
  $\Lambda=10^{-2}$, and (d) $\Lambda=1$ for $\alpha=1$ (green diamonds),
  $\alpha=2$ (magenta circles), $\alpha=3$ (red triangles), and $\alpha=4$
  (blue squares). Black lines show the corresponding theory curves,
  Eq.~(\ref{eq:alphaEntropyTheory}).}
\end{figure}

A bipartite system whose subsystems exhibit classical chaotic motion is considered here to compare against the universal entanglement dynamics results derived from random matrix theory. The knowledge of $\Lambda$ and its relation to the system dependent details is crucial for the comparison.
In case of a system whose subsystems are kicked rotors quantized on the unit torus, it is possible to analytically find $\Lambda$ as a function of system dependent details as shown in Refs.~\cite{Srivastava16,Tomsovic18c}. The Floquet unitary operator of the system has the form given by Eq.~(\ref{eq:GenericFloquet}) where the subsystem Floquet operator for one kicked rotor is
\begin{equation}
U_A = \exp[- i p_A^2/(2\hbar)] \exp(-i V_A/\hbar),
\end{equation}
with kicking potential given by
\begin{equation}
V_A = K_A \cos(2\pi q_A)/4\pi^2,
\end{equation}
where $K_A$ is the kicking strength. Similarly for subsystem $B$. The entangling operator is
\begin{equation}
U_{AB}(b) = \exp(-i b V_{AB}/\hbar),
\end{equation}
where the interaction potential is
\begin{equation}
V_{AB} = \frac{1}{4\pi^2} \cos[2\pi(q_A+q_B)].
\end{equation}
The angle variables $q_j$ is restricted to the interval $[0,1)$, and similarly for the momenta $p_j$. This restriction leads to a 4-dimensional torus phase space for the corresponding classical system \cite{Froeschle71,Lakshminarayan2001,Richter14}. The kicking strengths $(K_A,\, K_B) = (10,14),\,(18,22),\, \ldots$ with up to 20 realizations are chosen such that the classical dynamics is chaotic. The boundary conditions are chosen such that both time-reversal invariance and parity symmetry are broken. Thus the subsystem spectral fluctuations are approximately like those of the CUE. In addition we use $N=N_A = N_B = 100$ for the numerical computations. The transition parameter for the coupled kicked rotors is \cite{Srivastava16,Tomsovic18c,HerKieFriBae2019:p}
\begin{equation}
\Lambda_{\text{KR}} \simeq \frac{N^2}{4\pi^2} \left(1-J_0^2(Nb/2\pi) \right) \approx \frac{N^4 b^2}{32 \pi^4},
\end{equation}
where $J_0(\cdot)$ is the Bessel function of first kind
\cite[Eq.~10.2.2]{DLMF},
and the approximation is true when $Nb \ll 1$. In Fig.~\ref{fig:KR_Salpha}, the entanglement dynamics for various $\Lambda$-values of the coupled kicked rotors is shown against the theory given by Eq.~(\ref{eq:alphaEntropyTheory}).
Overall good agreement is found
with some small deviations for the von Neumann entropy
which are similar to those found for the
CUE case shown in Fig.~\ref{fig:CUE_Salpha}.

\section{Summary and outlook}
\label{sec:summary-and-outlook}

An analytic theory is given in this paper for the rate of entanglement production for a quenched system as a function of the interaction strength between chaotic subsystems.  In particular, all the expressions are given in terms of the universal transition parameter $\Lambda$.  It is shown that in the perturbative regime for an initial product of subsystem eigenstates (a so-called quench), the entanglement saturates at very small values proportional to $\sqrt{\Lambda}$.  Furthermore, in the same regime, once the appropriate time scale is properly identified and the entanglement entropies are scaled by their saturation value, there exists a single universal entropy production curve: for a given system size, the interaction strength determines $\Lambda$, which determines the time scale and saturation values, and there is no other dependence in the entropy production beyond that.  The universal curve has an overshoot, which is slightly more pronounced for the time reversal non-invariant case, and then it settles down to a saturation value.  As $\Lambda$ increases, the perturbation regime eventually breaks down, roughly for $\Lambda \gtrsim 10^{-2}$, as illustrated in Fig.~\ref{fig:UniversalCurvePlot}.

As for the full eigenstates of the interacting system \cite{Tomsovic18c}, it was also possible here to recursively embed the perturbation theory.  This enables a description of the full transition in entropy production behaviors as a function of subsystem interaction strength and size, the limiting behaviors being no entanglement entropy production for non-interacting systems, and for strongly interacting systems the production behavior seen for initially random product states.  The expressions are uniformly valid for all times and interaction strengths.  It also turns out that the initial entropy production rate is even independent of whether time reversal symmetry is preserved or not.

The present study also raises various interesting
questions to be addressed in the future:
the considered case of initial states given by direct products of subsystem eigenstates has the crucial property
that the automatic Schmidt decomposition holds,
which allows for a perturbative treatment.
If one considers instead,
for example, sums of such eigenstates or
direct products of subsystem random vectors, then a much faster entanglement
generation occurs, which requires a completely different
theoretical description.
Moreover, although not shown, the fluctuations of the entropies seen from one initial
state to another depend dramatically on whether it is subsystem eigenvectors or random states which are being considered.  This should also be reflected in the statistics
of Schmidt eigenvalues, which are expected to show
heavy-tailed distributions as was found before in the case of eigenstates.
Another interesting ensemble for the case of a dynamical system
like the coupled kicked rotors are coherent states
as initial states. There one will have an initial phase
for which the entanglement only grows very slowly
up to the Ehrenfest time beyond which a fast increase
of entanglement occurs.
Finally, bipartite many-body systems, like an interacting spin-chain,
should share many of the features
of the entanglement production demonstrated here,
and at the same time also allow for even more possibilities
of initial states.

\acknowledgments

We would like to thank Maximilian Kieler for useful discussions.
One of the authors (ST) gratefully acknowledges support for visits
to the Max-Planck-Institut f\"ur Physik komplexer Systeme.

\appendix
\section{Regularized Schmidt eigenvalues} \label{app:RegularizationDerivation}

In this section, the emergence of the regularized Schmidt eigenvalues in Eq.~(\ref{eq:RegularizedNormTimesMatrixElement}) is illustrated. The perturbation expression of the matrix element $S_{j'k',jk}$, relevant up to $\mathcal{O}(\epsilon^4)$, is
\begin{align}
S_{jk,j'k'} =
\begin{cases}
\frac{1}{\sqrt{\mathcal{N}_{jk}}} \qquad \qquad  \qquad \qquad \  jk = j'k' \\
\frac{1}{\sqrt{\mathcal{N}_{jk}}} \sum_{m=1}^3 \epsilon^m S^{(m)}_{jk,j'k'}  \quad jk \neq j'k',
\end{cases}
\end{align}
where $\mathcal{N}_{jk}$ is the normalization factor, and various corrections
read
\begin{subequations}
\begin{eqnarray}
S^{(1)}_{jk , j'k'} &=& \frac{V_{j'k',jk}}{\Delta\theta_{jk,j'k'}}, \\
S^{(2)}_{jk,j'k'} &=& \sum_{j''k'' \neq jk} \frac{V_{j'k',j''k''} V_{j''k'' ,jk}}{	\Delta\theta_{jk,j'k'}\Delta\theta_{jk,j''k''}}, \\
S^{(3)}_{jk,j'k'} &=& \sum_{j''k'',j'''k''' \neq jk}\frac{V_{j'k',j''k''} V_{j''k'',j'''k'''} V_{j'''k''',jk}}{\Delta\theta_{jk,j'k'} \Delta\theta_{jk,j''k''} \Delta\theta_{jk,j'''k'''} } \nonumber \\
&& -  \sum_{j''k'' \neq jk} \frac{|V_{j''k'',jk}|^2 V_{j'k',jk}}{\Delta\theta_{jk,j''k''} \Delta\theta_{jk,j'k'}^2} \label{eq:3rdOrderCorrection}.
\end{eqnarray}
\end{subequations}
Clearly the $S^{(1)}$ term involves only two levels, whereas each term in $S^{(2)}$ has three levels. In case of $S^{(3)}$, the first sum in Eq.~(\ref{eq:3rdOrderCorrection}) has four levels involved, but the second sum has a term involving only two levels when $j''k'' = j'k'$, i.e., $-|V_{j'k',jk}|^2 V_{j'k',jk}/\Delta \theta_{jk,j'k'}^3$, otherwise the terms have three levels. Substituting the perturbation expression for the matrix elements $S_{j'k',jk}$ in Eq.~(\ref{eq:LargestEigenvalueRaw}) and simplifying gives
\begin{align}
& \lambda_1(n;\epsilon) \nonumber \\
& = 1 - \frac{4}{\mathcal{N}_{jk}} \sum_{j'k' \neq jk} \frac{1}{\mathcal{N}_{j'k'}} \Bigg\{\epsilon^2 |S_{j'k',jk}^{(1)}|^2 \nonumber \\
& \quad\qquad + \epsilon^3 \big[ S_{j'k',jk}^{(1)} S_{j'k',jk}^{(2)*} + cc \big] \nonumber \\
&  \quad\qquad + \epsilon^4 \Big( |S_{j'k',jk}^{(2)}|^2 + \big[ S_{j'k',jk}^{(1)} S_{j'k',jk}^{(3)*} + cc \big] \Big) \Bigg\}  \nonumber \\
& \qquad \qquad \qquad \times \sin^2 \bigg( \frac{n \Delta \varphi_{j'k',jk}}{2} \bigg) \nonumber \\
& \quad - 4 \epsilon^4 \sum_{\substack{j'k' < j''k'' \\ \neq jk}} \frac{|S_{j'k',jk}^{(1)}|^2 |S_{j''k'',jk}^{(1)}|^2}{\mathcal{N}_{j'k'} \, \mathcal{N}_{j''k''}}
 \nonumber \\
& \qquad \qquad \qquad \times
 \sin^2 \bigg( \frac{n \Delta \varphi_{j'k',j''k''}}{2} \bigg).
\end{align}
The normalization factor to the relevant order is $1/\mathcal{N}_{jk} = 1 - \epsilon^2 \sum_{j''k'' \neq jk} |S_{j''k'',jk}|^2$. After separating the terms involving two levels up to $\mathcal{O}(\epsilon^4)$ one finds
\begin{align}
&\lambda_1(n;\epsilon) = 1 - 4 \sum_{j'k' \neq jk} \Big[ \epsilon^2 | S_{j'k',jk}^{(1)}|^2 - 4 \epsilon^4 | S_{j'k',jk}^{(1)}|^4 \Big] \nonumber \\
& \qquad \qquad \times \sin^2 \bigg( \frac{n \Delta \varphi_{j'k',j''k''}}{2} \bigg) + \mathcal{R}_{1,\text{HL}},
\end{align}
where all the higher level terms are grouped into $\mathcal{R}_{1,\text{HL}}$. The terms that appear in the square bracket above are the first two terms in the expansion of the regularized expression given in Eq.~(\ref{eq:RegularizedNormTimesMatrixElement}) obtained from the two-dimensional perturbation theory results,
\begin{equation}
\epsilon^2 | S_{j'k',jk}^{(1)}|^2 - 4 \epsilon^4 | S_{j'k',jk}^{(1)}|^4 + \ldots = \frac{\epsilon^2 | S_{j'k',jk}^{(1)}|^2}{1+ 4 \epsilon^2 | S_{j'k',jk}^{(1)}|^2}.
\end{equation}
For the other Schmidt eigenvalues ($l > 1$) a similar analysis can be shown.

\section{Computation of $C_{2}(\alpha;t)$}
         \label{app:C-2-alpha-t-derivation}

The first step to compute the $C_{2}(\alpha;t)$ integral Eq.~\eqref{eq:C2Integral}
is to use the identity $\sin^2(\theta/2) = (1-\cos\theta)/2$ and
then binomially expand
\begin{equation}
\left( 1- \cos\theta \right)^\alpha = \sum_{q=0}^\infty (-1)^q \binom{\alpha}{q} \cos^q\theta,
\end{equation}
and use the following Fourier series expansion
\begin{equation}
\cos^q\theta = \sum_{m=0}^q a_{qm} \cos(m\theta),
\end{equation}
where
\begin{align}
& a_{qm} = \binom{q}{\frac{q-m}{2}} \left[ \frac{1+(-1)^{q-m}}{2^q(1+\delta_{m,0})} \right] .
\end{align}
With this we get
\begin{equation}
\left( 1- \cos\theta \right)^\alpha = \sum_{q=0}^\infty \sum_{m=0}^q (-1)^q \binom{\alpha}{q} a_{qm} \cos(m\theta).
\end{equation}
Plugging in these results into Eq.~(\ref{eq:C2Integral}) gives
\begin{equation}
C_{2}(\alpha;t) = 2^\alpha \sum_{q=0}^\infty \sum_{m=0}^q (-1)^q \binom{\alpha}{q} a_{qm} f_{m}(\alpha;t),
\end{equation}
where
\begin{align}
& f_m(\alpha;t) = \int_{-\infty}^\infty \dd{z} \int_0^\infty \dd{w} \frac{w^\alpha \rho(w)}{(z^2+4w)^\alpha} \nonumber \\
&\qquad\qquad \qquad \times \cos(m t \sqrt{z^2+4w}\,).
\end{align}
In the above integral, the integrand is even in the $z$-variable, and combining it with the substitution
\begin{equation}
\sin^2 \theta = \frac{1}{2}\left( 1 - \frac{z}{\sqrt{z^2+4w}} \right),
\end{equation}
and simplifying gives
\begin{align}
& f_{m}(\alpha;t) = 2^{3-2\alpha}\int_0^{\pi/4} \dd{\theta}\int_0^\infty \dd{w} \sin^{2\alpha-2}(2 \theta) \nonumber \\
& \qquad\qquad \qquad \times \rho(w) \,\sqrt{w}\cos(\frac{2 m t \sqrt{w}}{\sin 2 \theta}).\label{eq:fmt_Integral}
\end{align}
In the next two subsections results specific to orthogonal and unitary ensembles are presented.

\subsection{Orthogonal case}
\label{app:C-2-alpha-t-derivation-COE}

In Eq.~(\ref{eq:fmt_Integral}), with $\rho(w)$ for the COE and integrating over $w$-variable gives
\begin{align}
& f_{m}(\alpha;t) = \frac{2^{3-2\alpha}}{\sqrt{\pi}} \int_0^{\pi/4} \dd{\theta} \bigg[ \sqrt{2}- 4 m t\csc(2\theta)  \nonumber \\
&\qquad \, \qquad \times F(\sqrt{2}\,mt\csc(2\theta)) \bigg] \sin^{2\alpha-2}(2\theta),
\end{align}
where $F(\cdot)$ is Dawson's integral \cite[Eq.~7.2.5]{DLMF}.
For $m = 0$ the above integral reduces to
\begin{align}
& f_0(\alpha;t) = \frac{2^{3/2-2\alpha} \Gamma(\alpha-1/2)}{\Gamma(\alpha)},
\end{align}
whereas for $m \neq 0$ the integral can then be represented in terms of special functions as
\begin{align}
& f_{m}(\alpha;t) = 2^{3-2\alpha}\bigg[ \frac{\Gamma(\alpha-1/2)}{\Gamma(\alpha) \sqrt{8}} + \frac{2^{\alpha-2} \pi (mt)^{2\alpha-1}}{\Gamma(\alpha)}  \nonumber \\
&\quad \qquad \qquad \times M(1/2,\alpha;-2 m^2 t^2) \sec(\pi \alpha) \nonumber \\
& \qquad \qquad - \frac{\sqrt{2}\,(mt)^2\Gamma(\alpha-3/2)}{\Gamma(\alpha-1)} \nonumber\\
&\qquad\qquad\quad \times \,_{2}F_{2}(1,2-\alpha\,;\,3/2,5/2-\alpha\,;\,-2 m^2 t^2) \bigg],
\end{align}
where $_{2}F_2(\cdot)$ is generalized hypergeometric function
\cite[Eq.~35.8.1]{DLMF}
and $M(\cdot)$ is the Kummer confluent hypergeometric function of the first kind
\cite[Eq.~13.2.2]{DLMF}.

\subsection{Unitary case}
\label{app:C-2-alpha-t-derivation-CUE}

Using $\rho(w)$ for the CUE case in Eq.~(\ref{eq:fmt_Integral}), and after integrating over $w$-variable, we get for $m= 0$
\begin{equation}
f_0(\alpha;t) = \frac{2^{3/2-2\alpha} \Gamma(\alpha-1/2)}{8 \, \Gamma(\alpha)},
\end{equation}
and for $m \neq 0$ we have
\begin{equation}
f_{m}(\alpha;t) = 2^{3-2\alpha} \left[ g_m(\alpha;t) - 2 m^2 t^2 g_m(\alpha-1;t) \right],
\end{equation}
where
\begin{equation}
g_m(\alpha;t) = \frac{\sqrt{\pi}}{2}\int_0^{\pi/4} \dd{\theta} \sin^{2\alpha-2}(2\theta) \exp(-m^2 t^2 \csc^2 2\theta).
\end{equation}
The above integral can be represented using special functions as
\begin{align}
& g_m(\alpha;t) = \frac{\pi^{3/2} }{8\cos(\pi \alpha)} \bigg[(m t)^{2\alpha-1}\frac{M(1/2,\alpha+1/2,- m^2 t^2)}{\Gamma(\alpha+1/2)} \nonumber \\
& \quad \qquad \qquad - \sqrt{\pi} \frac{M(1-\alpha,3/2-\alpha,-4\pi^2 m^2 t^2)}{\Gamma(\alpha) \Gamma(3/2-\alpha)} \bigg] .
\end{align}

\subsection{von Neumann entropy}

For the critical value $\alpha=1$ Eq.~(\ref{eq:GenericHCTEntropy}) gives
the von Neumann entropy as
\begin{equation}
\overline{S_1(t;\Lambda)} = \sqrt{\Lambda} \,\lim_{\alpha\rightarrow 1} \pdv{}{\alpha} C(\alpha;t). \label{eq:S1}
\end{equation}
The derivative in Eq.~(\ref{eq:S1}) gives
\begin{align}
& \lim_{\alpha \rightarrow 1} \pdv{C(\alpha;t)}{\alpha} = \lim_{\alpha\rightarrow 1} \sum_{p=1}^\infty (-1)^{p+1} C_{2}(p;t) \pdv{\alpha} \binom{\alpha}{p} \nonumber \\
&\qquad \qquad \qquad \qquad - \,  \lim_{\alpha \rightarrow 1} \pdv{\alpha} C_{2}(\alpha;t). \label{eq:diffC1exp}
\end{align}
It can be shown that for $p \in \mathbb{Z}^+$
\begin{equation}
\lim_{\alpha \rightarrow 1} \pdv{\alpha} \binom{\alpha}{p} =
\begin{cases}
1, \qquad \qquad \qquad \,\text{for} \, p = 1, \nonumber \\
\frac{(-1)^p}{p(p-1)}, \qquad \qquad \text{for}\, p > 1.
\end{cases}
\end{equation}
Using this in Eq.~\eqref{eq:diffC1exp} gives
\begin{align}
& \lim_{\alpha \rightarrow 1} \pdv{C(\alpha;t)}{\alpha} = C_2(1;t) - \sum_{p=2}^\infty \frac{C_2(p;t)}{p(p-1)} \nonumber \\
& \qquad \qquad \qquad \quad - \lim_{\alpha \rightarrow 1} \pdv{\alpha} C_{2}(\alpha;t). \label{eq:diffC1expSimplified}
\end{align}
Now it amounts to computing $\pdv*{C_2(\alpha;t)}{\alpha}$ in the limit $\alpha \rightarrow 1$,
\begin{align}
& \lim_{\alpha \rightarrow 1} \pdv{\alpha} C_{2}(\alpha;t) = C_{2}(1;t)\, \ln 2 - 2 f_{1}(1;t) \nonumber \\
& \qquad \qquad \qquad \qquad + 2 \pdv{f_{0}(\alpha;t)}{\alpha}\bigg|_{\alpha =1} - 2 \pdv{f_{1}(\alpha;t)}{\alpha}\bigg|_{\alpha =1} \nonumber \\
& \qquad\qquad \qquad \qquad + 2 \sum_{q=2}^\infty \sum_{m=0}^q \frac{a_{qm}\,f_{m}(1;t)}{q(q-1)}.
\end{align}

In the above, the derivatives are
\begin{equation}
\pdv{f_{0}(\alpha;t)}{\alpha}\bigg|_{\alpha =1} =
\begin{cases}
-\sqrt{8 \pi} \, \ln 2, \qquad \text{for COE,} \\
- \pi^{3/2} \, \ln 2, \qquad \text{for CUE}
\end{cases}
\end{equation}
and
\begin{equation}
\pdv{f_{1}(\alpha;t)}{\alpha}\bigg|_{\alpha =1} = - f_{1}(1;t) \, \ln 4 + h(t).
\end{equation}
The last term in the above equation is an integral, and is different for
the COE  and the CUE.
For the COE case one gets
\begin{align}
&h(t) = \frac{4}{\sqrt{\pi}} \int_0^{\pi/4} \dd{\theta} \big[\sqrt{2} -4 t \csc(2\theta) F(\sqrt{2} t \csc2\theta)\big]  \nonumber \\
& \qquad \qquad \qquad \times \, \ln \sin 2\theta,
\end{align}
and can be represented using special functions,
\begin{align}
& h(t) = \sqrt{2\pi} \big[4 t^2 - \ln 2 \big] - \pi t \pdv{\alpha} M(1/2,\alpha,-2 t^2) \bigg|_{\alpha=1} \nonumber \\
& \qquad \quad - \frac{32 t^4 \sqrt{2 \pi}}{9} {}_2F_2(1,2;5/2,5/2;-2 t^2) \nonumber \\
& \qquad \quad - \ue^{-t^2} \pi \, t \, I_0(t^2) \big[\gamma + \ln 2 + 2 \ln t \big].
\end{align}
Similarly for CUE case one gets
\begin{align}
& h(t) = 2\sqrt{\pi} \int_0^{\pi/4} \dd{\theta}  \ue^{-t^2\csc^2 2\theta}\ln(\sin 2 \theta)(1-2 t^2 \csc^2 2\theta), \nonumber \\
\end{align}
which again can be represented using special functions,
\begin{align}
& h(t) = \frac{\pi^{3/2}}{4}\bigg[ 2 t^2 {}_2F_2(1,1;3/2,2;-t^2) -\ln 4 - 2 \text{erf}(t) \ln t \nonumber \\
& \qquad \quad - \frac{2 \, t}{\sqrt{\pi}}\, \pdv{\alpha} M(1/2,\alpha,-t^2)\big|_{\alpha=3/2} \bigg] - \frac{\pi \, t \, \ue^{-t^2}}{2}\bigg[ \gamma - \nonumber \\
& \qquad \quad - \pi \,\text{erfi}(t) + \ln 4 + 2 \ln t + \pdv{\alpha} M(\alpha,1/2,t^2)\big|_{\alpha=0} \bigg], \nonumber \\
\end{align}
where $\text{erfi}(z) = \text{erf}(i z)/i$ is the imaginary error function.

\bibliographystyle{cpg_unsrt_title_for_phys_rev}

\bibliography{abbrevs,general_ref,classicalchaos,extracted,extreme,furtherones,quantumchaos,manybody,references,rmtmodify}
\end{document}